\documentstyle[amssymb,preprint,eqsecnum,aps]{revtex}

\def\a{\alpha}

\def\y{\eta}

\newcommand{\be}{\begin{equation}}
\newcommand{\ee}{\end{equation}}
\newcommand{\bn}{\begin{eqnarray}}
\newcommand{\en}{\end{eqnarray}}

\newcommand{\ba}{\begin{array}}
\newcommand{\ea}{\end{array}}

\begin{document}
\draft
\title{{\Large Quantization of Point-Like Particles and Consistent Relativistic
Quantum Mechanics}}
\author{S.P. Gavrilov\thanks{%
Universidade Federal de Sergipe, Brasil; on leave from Tomsk Pedagogical
University, Russia; present e-mail: gavrilov@ufs.br} and D.M. Gitman\thanks{%
e-mail: gitman@fma.if.usp.br}}
\address{Instituto de F\'{\i}sica, Universidade de S\~ao Paulo\\
P.O. Box 66318, 05315-970 S\~ao Paulo, SP, Brasil}
\date{\today}
\maketitle

\begin{abstract}
We revise the problem of the quantization of relativistic particle models
(spinless and spinning), presenting a modified consistent canonical scheme.
One of the main point of the modification is related to a principally new
realization of the Hilbert space. It allows one not only to include
arbitrary backgrounds in the consideration but to get in course of the
quantization a consistent relativistic quantum mechanics, which reproduces
literally the behavior of the one-particle sector of the corresponding
quantum field. In particular, in a physical sector of the Hilbert space a
complete positive spectrum of energies of relativistic particles and
antiparticles is reproduced, and all state vectors have only positive norms.
\end{abstract}


\newpage

\section{Introduction}

Already for a long time there exists a definite interest in studying
(construction and quantization) of classical and pseudoclassical models of
relativistic particles (RP) of different kinds. There are various reasons
for that. One may mention, for example, a widely spread explanation that RP
is a prototype string and with that simple example one can study many of the
problems related to string quantization. However, we believe that a more
profound motivation is stipulated by a desire to understand better basic
principles of the quantization theory and to prove that there exist a
consistent classical and quantum descriptions at least for noninteracting
(between each other) RP of different kinds (with different masses, spins,
and in different space-time dimensions), moving in external backgrounds. The
problem may be considered as a supplementary one to the problem of
relativistic wave equation construction for the particles of different
kinds. Indeed, quantizing a classical or pseudoclassical model of RP we
aspire to reproduce a quantum mechanics, which in a sense is based on the
corresponding relativistic wave equation. And here it is necessary to
formulate more precisely the aim of the quantization problem. Indeed, there
is a common opinion that the construction of a consistent relativistic
quantum mechanics on the base of the relativistic wave equations meets
well-known difficulties related to the existence of infinite number of
negative energy levels (energy levels, which correspond to antiparticles
appear with negative sign in the spectrum), to the existence of negative
vector norms, and difficulties related to localized state construction
(position operator problem), part of which may be only solved in the
second-quantized theory, see for example \cite
{FolWo50,FesVi58,Schwe61,GroSu72,Grein97,Weinb95}. In this relation one
ought to mention some attempts to to construct relativistic wave equations
for the wave functions, which realize infinite-dimensional representations
of the Lorentz group \cite{Major32,Dirac71}. Thus, the quantization problem
under consideration may be formulated with different degrees of claim. The
simplest and most widely used approach is to apply some convenient in the
concrete case (but not always the most convinced and well-grounded) scheme
of quantization in the given case to arrive in a way to a corresponding
relativistic wave equation, without any attempt to demonstrate that a
consistent quantum mechanics was constructed, due to the above mentioned
belief that it cannot be done. To our mind the aim has to be more ambitious:
namely, in the course of a first quantization of a RP model one has to try
to construct a relativistic quantum mechanics consistent to the same extent
to which a one-particle description is possible in the frame of the
corresponding quantum field theory (in the frame of the second quantized
theory). We will demonstrate in the present article a possible way of
realizing this with the example of the quantization problem of a spinless
and spinning charged RP moving in arbitrary external electromagnetic and
gravitational fields.

First of all we have to define more precisely what we mean by convincing and
well-grounded quantization. To our mind it has to be a consistent general
scheme, but not some leading considerations, which allow one to predict in a
way some basic aspects of a corresponding quantum theory of the classical
model under consideration. For a consistent scheme of this nature one may
refer, for example, to the canonical quantization of gauge theories, in
which the physical sector may be selected already on the classical level by
means of a gauge fixing, and the state space may be constructed and analyzed
in detail. That may be also any equivalent to the canonical quantization
scheme, which allows one to achieve the same final result. An alternative
and frequently used method of Dirac quantization, in which the gauge
conditions are not applied on the classical level, and first-class
constraints are used as operators to select the physical sector in the state
space, contains some essential intrinsic contradictions, in particular, one
cannot formulate a consistent prescription to construct the appropriate
Hilbert space in this case. Besides, there is no general proof of the
equivalence of this method to the canonical quantization. All that does not
allow one to consider this quantization scheme as a consistent one in the
above mentioned sense. One ought to say that this method is rather popular
due to its simplicity and due to the possibility to sometimes quickly reach
a desired result. In particular, from the point of view of this method the
problem of quantization, for example of scalar RP looks, in a sense,
trivial. Indeed, the first-class constraint $p^{2}=m^{2}$ reproduces in this
scheme immediately something which looks like Klein-Gordon equation. In
spite of the fact that still nothing has been said how a consistent quantum
mechanics may appear starting with that point, sometimes they accept it as a
final solution of the problem. In connection with this we would like to
repeat once again that the problem, as we see it, is not to ``derive'' in
any way the Klein-Gordon or Dirac equations. The question is: May or not a
consistent relativistic quantum mechanics be reproduced in the course of
honest application of a well developed general scheme of a quantization, to
some classical or pseudoclassical models of RP? Under the consistent
relativistic quantum mechanics we mean a reduction of the quantum field
theory of a corresponding field (scalar, spinor, etc.) to the one particle
sector, if such a reduction may be done. The latter is possible if the
interaction of the given quantum field with external backgrounds does not
lead to a particle creation.

What was done before to solve such formulated problem? Which kind of
difficulties one meets here, and how the present work may contribute to
progress in this direction?

Usually the above mentioned models of RP are formulated in covariant and
reparametrization invariant form. Due to the latter invariance, which is, in
fact, a gauge invariance, one meets here all the problems related to the
quantization of such systems, e.g. zero-Hamiltonian phenomenon and the time
problem, which are crucial, for example, also for the quantization of such
important reparametrization invariant theory as general relativity. Besides,
the problem of spinning degree of freedom description turns out to be
nontrivial in RP models. Here there are two competing approaches, one which
uses Grassmann variables for spin description, and gives rise to the
pseudoclassical mechanics, and another one, which uses variables from a
compact bosonic manifold. Both approaches have their own problems related,
in particular, to higher spin description and introduction for such spins an
interaction to external backgrounds. We do not touch here the problem of
path integral quantization of relativistic particles. Readers interested in
that question may look, for example, up the articles \cite{path}.

One of possible approach to the canonical quantization of the relativistic
particle (spinless and spinning) was presented in the papers \cite
{GitTy90a,GitTy90b} on the base of a special gauge, which fixes the
reparametrization gauge freedom. It was shown how the Klein Gordon and Dirac
equations appear in the course of the quantization from the corresponding
Schr\"{o}dinger equations. However, only a restricted class of external
backgrounds (namely, constant magnetic field) was considered. One may see
that the above quantum theory does not obey all symmetries of the
corresponding classical model. Besides, an analysis of the equivalence
between the quantum mechanics constructed in course of the quantization and
the one-particle sector of the corresponding field theory was not done in
all details. Moreover, from the point of view of the results of the present
work, one can see that such an equivalence is not complete. Thus, the
question: whether a consistent quantum mechanics in the above mentioned
sense is constructed, remain. Attempts to generalize the consideration to
arbitrary external electromagnetic background \cite{GriGr95} have met some
difficulties (even Klein-Gordon and Dirac equations were not reproduced in
course of the quantization), which look even more complicated in the case of
a RP moving in curved space-time (in an external gravitational field). As to
the latter problem, it is enough to mention that even more simple
corresponding nonrelativistic problem (canonical quantization of a particle
in curved three-dimensional space, which has a long story \cite{DeWit52})
attracts attention to the present day and shows different points of view on
its solution \cite{DeWit52,Marin80,OgaFuK90,Saa97}. The relativistic
problem, which naturally absorbs all known difficulties of its
nonrelativistic analog, is essentially more rich and complicated due to its
gauge nature (reparametrization invariance). If the external gravitational
field is arbitrary, then the problem can not be solved (even in the
restricted sense to reproduce only the Klein-Gordon and Dirac equations) by
complete analogy with the flat space case in an external constant magnetic
field \cite{GitTy90b}, how it was done in \cite{Saa96} for the static
space-time. However, namely the general case is interesting from the
principle point of view. It turns out that the whole scheme of quantization,
which was used in \cite{GitTy90a,GitTy90b} and repeated then in numerous
works, has to be changed essentially to make it possible to include
arbitrary external backgrounds (electromagnetic or gravitational) in the
consideration and maintain all classical symmetries on the quantum level.
Such a modified scheme of the canonical quantization of RP is described in
the present article first in detail on the example of a spinless charged
particle moving in arbitrary external electromagnetic and gravitational
fields, and then it is applied already briefly to the spinning particle case.

One of the main point of the modification is related to a principally new
realization of the Hilbert space. It has allowed one not only to include
arbitrary backgrounds in the consideration but to solve the problem
completely, namely to get in course of the quantization the consistent
relativistic quantum mechanics, which reproduces literally the behavior of
the one-particle sector of the quantum field (in external backgrounds, which
do not create particles from vacuum). In particular, in a physical sector of
the Hilbert space complete positive spectrum of energies of relativistic
particles and antiparticles is reproduced, and corresponding state vectors
have only positive norms.

The article is organized in the following way: In Sect.II we present a
detailed Hamiltonian analysis of the theory of a classical relativistic
particle with a reparametrization invariant action in external
electromagnetic and gravitational backgrounds. We focus our attention on the
selection of physical degrees of freedom and on the adequate gauge fixing.
Due to the fact that after our gauge fixing we remain with time dependent
constraint system, a method of treatment and quantization of such systems is
briefly discussed in the end of the Section. In Sect.III we proceed with
canonical first quantization procedure. Here we discuss in details Hilbert
space construction, realization of all physical operators in this space, and
in the end of the section we reformulate the evolution of the system under
consideration in terms of a physical time. In Sect.IV we demonstrate a full
equivalence of the quantum mechanics constructed to the dynamics of
one-particle sector of the corresponding field theory in backgrounds, which
do not create particles from vacuum. In the Sect. V we generalize
consideration to the spinning particle case. To make the consideration
complete we present also Dirac quantization scheme both in scalar and
spinning case. Treating the spinless case we use widely results of brief
consideration of the quantum field theory of scalar field in the
electromagnetic and gravitational backgrounds, which are presented in the
Appendix to the article.

\section{Classical spinless relativistic particle}

The classical theory of a relativistic charged spinless particle placed in $%
3+1$-dimensional Riemannian space-time (with coordinates $x=(x^{\mu
})=(x^{0},x^{i})=(x^{0},{\bf x})$, and a metric tensor $g_{\mu \nu
}(x),\;\mu =0,1,2,3,\;\;i=1,2,3$), and interacting with an external
electromagnetic field, may be described by a reparametrization invariant
action 
\begin{equation}
S=\int_{0}^{1}Ld\tau \;,\;\;L=-m\sqrt{\dot{x}^{\mu }g_{\mu \nu }(x)\dot{x}%
^{\nu }}-q\dot{x}^{\mu }g_{\mu \nu }(x)A^{\nu }(x)=-m\sqrt{\dot{x}^{2}}-q%
\dot{x}A\;,  \label{1}
\end{equation}
where $\dot{x}^{\mu }=dx^{\mu }/d\tau $; $\tau $ is a real evolution
parameter, which plays the role of time in the problem under consideration; $%
q$ is an algebraic charge of the particle; and $A^{\mu }(x)$ are potentials
of an external electromagnetic field. The action (\ref{1}) is invariant
under reparametrizations $x^{m}(\tau )\rightarrow x^{^{\prime }\mu }(\tau
)=x^{m}(f(\tau ))$, where $f(\tau )$ is an arbitrary function subjected only
the following conditions: $\dot{f}(\tau )>0,\;f(0)=0,\;f(1)=1$. The
reparametrizations may be interpreted as gauge transformations whose
infinitesimal form is $\delta x^{\mu }(\tau )=\dot{x}^{\mu }(\tau )\epsilon
(\tau ),\;$ where $\epsilon (\tau )$ is $\tau $-dependent gauge parameter.

For the purposes of the quantization it is preferable to select a reference
frame, which admits a time synchronization over all space. Such a reference
frame corresponds to a special gauge $g_{0i}=0$ of the metric\footnote{%
In such a gauge $g^{00}=g_{00}^{-1},\;g^{ik}g_{kj}=\delta _{j}^{i}\,.$}. It
is called (at $g_{00}=1$) synchronous reference frame according to \cite
{LanLi2}, or corresponds to Gaussian coordinates according to \cite{Synge60}%
. Such a reference frame exists always for any real space-time.

Our aim is the canonical quantization, thus, we need first a detailed
Hamiltonian analysis of the problem. Let us denote via $p_{\mu }$ canonical
momenta conjugated to the coordinates $x^{\mu }$, 
\begin{equation}
p_{\mu }=\frac{\partial L}{\partial \dot{x}^{\mu }}=-\frac{mg_{\mu \nu }\dot{%
x}^{\nu }}{\sqrt{\dot{x}^{2}}}-qA_{\mu }\;,\;\;A_{\mu }=g_{\mu \nu }A^{\nu
}\;.  \label{3}
\end{equation}
Let us introduce an important for the further consideration discrete
quantity $\zeta =\pm 1$, which is defined in the phase space, 
\begin{equation}
\zeta =-{\rm sign}\left[ p_{0}+qA_{0}\right] \;.  \label{5}
\end{equation}
Due to the fact that $g_{00}>0$ an important relation follows from the Eq. (%
\ref{3}) at $\mu =0$, 
\begin{equation}
{\rm sign}(\dot{x}^{0})=\zeta \;.  \label{6}
\end{equation}
It follows also from (\ref{3}) that there exists a primary constraint 
\begin{equation}
\Phi _{1}^{\prime }=\left[ p_{\mu }+qA_{\mu }\right] g^{\mu \nu }\left[
p_{\nu }+qA_{\nu }\right] -m^{2}=0\;,\;\;g^{\alpha \nu }g_{\nu \beta
}=\delta _{\beta }^{\alpha }\;.  \label{4}
\end{equation}
On the other hand, it is clear that the relation (\ref{4}) is, in fact, a
constraint on the modulus of $p_{0}+qA_{0}$ only, 
\begin{equation}
|p_{0}+qA_{0}|=\omega \,,\;\;\omega =\sqrt{g_{00}\left\{ m^{2}-\left[
p_{k}+qA_{k}\right] g^{kj}\left[ p_{j}+qA_{j}\right] \right\} }\;.  \label{8}
\end{equation}
Taking into account (\ref{5}), we may write an equivalent to (\ref{4})
constraint in the following linearized in $p_{0}$ form 
\begin{equation}
\Phi _{1}=p_{0}+qA_{0}+\zeta \omega =0\;.  \label{9}
\end{equation}
Indeed, it is easy to see that $\Phi _{1}^{\prime }$ is a combination of the
constraint (\ref{9}), $\Phi _{1}^{\prime }=g^{00}\left[ -2\zeta \omega \Phi
_{1}+\left( \Phi _{1}\right) ^{2}\right] \,.$ Further we are going to work
with the constraint $\Phi _{1},$ in particular, one see explicitly that it
imposes no restrictions on $\ \zeta .$ \ That is especially important for
our consideration.

To construct the total Hamiltonian in a theory with constraints we have to
identify the primary-expressible velocities and primary-inexpressible ones 
\cite{GitTy90}. The role of the former velocities are playing here ${\rm sign%
}(\dot{x}^{0})$ and $\dot{x}^{i}$. Indeed, the first quantity is expressed
via the phase space variables (see (\ref{6})). Besides, it follows from the
equation (\ref{3}) that 
\begin{equation}
\sqrt{\dot{x}^{2}}=mg_{00}\omega ^{-1}\lambda \;,\;\;\dot{x}%
^{i}=-g_{00}\omega ^{-1}\left[ p_{k}+qA_{k}\right] \lambda
g^{ki}\;,\;\;\lambda =|\dot{x}^{0}|\;.  \label{10}
\end{equation}
Thus, we may regard $\dot{x}^{i}$ as primary-expressible velocity as well,
and $\lambda $ as primary-inexpressible velocity. We may expect that the
latter quantity will appear as a Lagrange multiplier in the total
Hamiltonian $H^{(1)}$ of the theory. Indeed, constructing such a Hamiltonian
according to the standard procedure \cite{GitTy90,Dirac64,HenTe92}, we get 
\begin{equation}
H^{(1)}=\left. \left( p_{\mu }\dot{x}^{\mu }-L\right) \right| _{\dot{x}^{\mu
}=f(x,p,\lambda )}=\zeta \lambda \Phi _{1}\;.  \label{11}
\end{equation}
It vanishes on the constraint surface in accordance with the
reparametrization invariance nature of the formulation \cite
{HenTe92,Bergm42,FulGiT98,FulGiT99}. One can make sure that the equations 
\begin{equation}
\dot{\eta}=\{\eta ,H^{(1)}\},\;\;\Phi _{1}=0,\;\;\lambda >0,\;\;\eta
=(x^{\mu },p_{\mu })\;,  \label{12}
\end{equation}
are equivalent to the Lagrangian equations of motion\footnote{%
Here and in what follows the Poisson brackets are defined as \cite{GitTy90} 
\[
\left\{ {\cal F},{\cal G}\right\} =\frac{\partial _{r}{\cal F}}{\partial
q^{a}}\frac{\partial _{l}{\cal G}}{\partial p_{a}}-\left( -1\right) ^{P_{%
{\cal F}}P_{{\cal G}}}\frac{\partial _{r}{\cal G}}{\partial q^{a}}\frac{%
\partial _{l}{\cal F}}{\partial p_{a}}\;, 
\]
where $P_{{\cal F}}$ and $P_{{\cal G}}$ are Grassmann parities of ${\cal F}%
\, $and $\,{\cal G}$ respectively.} (with account taken of the definition of
the momenta (\ref{3})). Equations (\ref{12}) are equations of motion of a
Hamiltonian theory with a primary constraint (\ref{9}). In these equations $%
\lambda $ is an undetermined Lagrange multiplier, about which part of the
information ($\lambda >0$) is already available (the latter condition is
necessary to provide the above mentioned equivalence).

The consistency condition ($\dot{\Phi}_{1}=0$) for the constraint (\ref{9})
does not lead to any new secondary constraints and $\lambda $ is no longer
defined. Thus, (\ref{9}) is a first-class constraint. What kind of gauge
fixation one can chose to transform the theory to second-class constraint
type? Let us consider, for example, the case of a neutral ($q=0$) particle.
In this case the action (\ref{1}) is invariant under the time inversion $%
\tau \rightarrow -\tau $. Since the gauge symmetry in the case under
consideration is related to the invariance of the action under the changes
of the variables $\tau $, there appear two possibilities: namely, to include
or not to include the above discrete symmetry in the gauge group together
with the continuous reparametrizations. Let us study the former possibility
and include the time inversion in the gauge group. Then the gauge conditions
have to fix the gauge freedom, which corresponds to both kind of symmetries,
namely, to fix the variable $\lambda =|\dot{x}^{0}|$, which is related to
the reparametrizations, and to fix the variable $\zeta ={\rm \ sign}\,\dot{x}%
^{0}$, which is related to the time inversion. To this end we may select a
supplementary condition (the chronological gauge) of the form $\Phi
_{2}=x^{0}-\tau =0\;.$ The consistency equation $\dot{\Phi}_{2}=0$ leads on
the constraint surface to the relation $\partial _{\tau }\Phi _{2}+\{\Phi
_{2},H^{(1)}\}=-1+\lambda \zeta =0~,$ which results in $\zeta \lambda =1$.
Remembering that $\lambda \geq 0$, we get $\zeta =1,\;\lambda =1$. Suppose
we do not include the time inversion in the gauge group. That is especially
natural when $q\neq 0$,\ $A_{\mu }\neq 0$, since in this case the time
inversion is not anymore a symmetry of the action. Under the above
supposition the above suplimentary condition is not anymore a gauge, it
fixes not only the reparametrization gauge freedom (fixes $\lambda $) but it
fixes also the variable $\zeta $, which is now physical. A possible gauge
condition, which fixes only $\lambda $, has the form \cite{GitTy90a,GitTy90b}%
: 
\begin{equation}
\Phi _{2}=x^{0}-\zeta \tau =0~.  \label{15}
\end{equation}
Indeed, the consistency condition $\dot{\Phi}_{2}=0$ leads to the equation $%
\partial _{\tau }\Phi _{2}+\{\Phi _{2},H^{(1)}\}=-\zeta +\lambda \zeta =0~,$
which fixes $\lambda =1$ and retains $\zeta $ as a physical variable. To
make more clear the meaning of the discrete variable $\zeta =\pm 1$ let us
study the equations of motion (\ref{12}) in the gauge (\ref{15}). Selecting
for simplicity the flat space case ($g_{\mu \nu }=\eta _{\mu \nu }={\rm \
diag}(1,-1,\dots ,-1)$), we can see that these equations may be written in
the following form: 
\begin{eqnarray}
&&\frac{d{\cal P}_{i}^{kin}}{d(\zeta \tau )}=(\zeta q)\left[ F_{0i}+F_{ji}%
\frac{dx^{j}}{d(\zeta \tau )}\right] \,,\;\;F_{\mu \nu }=\partial _{\mu
}A_{\nu }-\partial _{\nu }A_{\mu }\;,  \nonumber \\
&&\frac{dx^{i}}{d(\zeta \tau )}=\frac{{\cal P}_{i}^{kin}}{\sqrt{m^{2}+({\cal %
P}_{kin}^{i})^{2}}}\,,\;\;\;\frac{d\zeta }{d(\zeta \tau )}=0\,,\;\;\;{\cal P}%
_{i}^{kin}=\zeta p_{i}+(\zeta q)A_{i}\;.  \label{17}
\end{eqnarray}
It is natural now to interpret $\zeta \tau =x^{0}$ as a physical time%
\footnote{%
In a sense we return into the consideration the initial variable $x^{0}$,
which was gauged out by means of a gauge condition. However, now $x^{0}$ has
the status of an evolution parameter but not a dynamical variable.}, $\zeta
p_{i}={\cal P}_{i}$ as a physical momentum, and $\frac{dx^{j}}{d(\zeta \tau )%
}=\frac{dx^{j}}{dx^{0}}=v^{j}$ as a physical three-velocity. Then ${\cal P}%
_{i}^{kin}={\cal P}_{i}+(\zeta q)A_{i}$ is the kinetic momentum of a
particle with the charge $\zeta q$. In terms of such quantities the
equations (\ref{17}) take the form: 
\begin{equation}
\frac{d{\mbox{\boldmath${\cal P}$\unboldmath}}_{kin}}{dx^{0}}=(\zeta
q)\left\{ {\bf E}+[{\bf v},{\bf H}]\right\} \,,\;{\bf v}=\frac{{%
\mbox{\boldmath${\cal P}$\unboldmath}}_{kin}}{\sqrt{m^{2}+{%
\mbox{\boldmath${\cal P}$\unboldmath}}_{kin}^{2}}},\;\;\frac{d\zeta }{dx^{0}}%
=0,\;\;\zeta =\pm 1\;,  \label{18}
\end{equation}
where ${\bf E}$ and ${\bf H}$ are electric and magnetic fields respectively
and ${\bf v}=(v^{i}),\;{\mbox{\boldmath${\cal P}$\unboldmath}}^{kin}=\left( 
{\cal P}_{i}^{kin}\right) $. Equations (\ref{18}) are well recognized
classical relativistic equations of motion for a charge $\zeta q$ moving in
an external electromagnetic field \cite{LanLi2}. Now we may conclude that
trajectories with $\zeta =+1$ correspond to the charge $q$, while those with 
$\zeta =-1$ correspond to the charge $-q$ (that was first pointed out in 
\cite{GitTy90a,GitTy90b}). The sign of the charge $\zeta $ is a conserved
quantity in the theory. This interpretation remains also valid in the
presence of an arbitrary gravitational background. Thus, the theory with the
action (\ref{1}) describes states with both sign of the electric charge.
This doubling of the state space on classical level (due to the existence of
the variable $\zeta $) naturally appears also on the quantum level, how it
will be demonstrated below, and is decisive for the construction of a
consistent relativistic quantum mechanics.

The set of the constraints $\Phi _a=0,\;a=1,2$ is now second-class. However,
it depends explicitly on the time (namely $\Phi_2 $ does). In this case an
usual canonical quantization by means of Dirac brackets has to be modified
(see for details \cite{GitTy90}). In the case of a particle in a flat space,
moving in an magnetic field with time independent potentials \cite
{GitTy90a,GitTy90b}, or in completely similar case of a particle in static
space-time \cite{Saa96}, it is possible to make explicitly a simple
canonical transformation, which transforms the constraint surface to a time
independent form, and then proceed to the usual scheme of the canonical
quantization by means of Dirac brackets. The above mentioned canonical
transformation depends explicitly on time, thus, a new effective
non-vanishing on the constraint surface Hamiltonian appears. In the case
under consideration, with arbitrary gravitational and electromagnetic
backgrounds, to find such a canonical transformation seems to be a difficult
task. Nevertheless, the problem of the canonical quantization may be solved
on the base of the approach to non-stationary second-class constraints
developed in \cite{GitTy90} (similar results were obtained by a geometrical
approach in \cite{EvaTu93}). Below we present such an approach, which allows
one to treat easily the backgrounds of general form and at the same time
clarifies some ambiguities hidden in the scheme of quantization of the
original papers \cite{GitTy90a,GitTy90b}, which was used in many following
publications devoted to the canonical quantization of the classical and
pseudoclassical models (see \cite{Gitma96} and Ref. therein). First, we
recall briefly the treatment of Ref. \cite{GitTy90} for systems with
non-stationary second-class constraints.

Consider a theory with second-class constraints $\Phi _{a}(\eta ,t)=0$
(where $\eta =(x^{i},\pi _{i})$ are canonical variables), which may
explicitly depend on time $t$. Then the equations of motion for such a
system may be written by means of the Dirac brackets, if one formally
introduces a momentum $\epsilon $ conjugated to the time $t$, and defines
the Poisson brackets in the extended phase space of canonical variables $%
(\eta ;t,\epsilon )$, 
\begin{equation}
\dot{\eta}=\{\eta ,H+\epsilon \}_{D(\Phi )},\;\;\;\Phi (\eta ,t)=0~,
\label{19}
\end{equation}
where H is the Hamiltonian of the system, and $\{A,B\}_{D(\phi )}$ is the
notation for the Dirac bracket with respect to the system of second--class
constraints $\phi $. The Poisson brackets, wherever encountered, are
henceforth understood as ones in the above mentioned extended phase space.
The quantization procedure in Heisenberg picture can be formulated in that
case as follows. The variables $\eta $ of the theory are assigned the
operators $\hat{\eta}$, which satisfy the following equations and
commutation relations\footnote{%
In fact, the commutator here is understood as a generalized one, it is a
commutator in case if one or both operators have Grassmann even parities,
and it is an anticommutator if both operators have Grassmann odd parities.} 
\begin{equation}
\dot{\hat{\eta}}=\left. \{\eta ,H+\epsilon \}_{D(\Phi )}\right| _{\eta =\hat{%
\eta}}\,,\;\;[\hat{\eta},\hat{\eta}^{\prime }]=i\left. \{\eta ,\eta ^{\prime
}\}_{D(\Phi )}\right| _{\eta =\hat{\eta}}~\;\;\Phi (\hat{\eta},t)=0\;.
\label{20}
\end{equation}
The total time evolution is controlled only by the first set of the
equations (\ref{20}) since the state vectors do not depend on time in the
Heisenberg picture. In the general case such an evolution is not unitary.
Suppose, however, that a part of the set of second-class constraints
consists of supplementary gauge conditions, the choice of which is in our
hands. In this case one may try to select these gauge conditions in a
special form to obtain an unitary evolution. The evolution is unitary if
there exists an effective Hamiltonian $H_{eff}(\eta )$ in the initial phase
space of the variables $\eta $ so that the equations of motion (\ref{19})
may be written as follows 
\begin{equation}
\dot{\eta}=\{\eta ,H+\epsilon \}_{D(\Phi )}=\{\eta ,H_{eff}\}_{D(\Phi
)}\,,\;\ \Phi (\eta ,t)=0\;.  \label{21}
\end{equation}
In this case, (due to the commutation relations (\ref{20})) the quantum
operators $\hat{\eta}$ obey the equations (we disregard here problems
connected with operator ordering) 
\begin{equation}
\dot{\hat{\eta}}=-i[\hat{\eta},\hat{H}]\,,\;\;\hat{H}=H_{eff}(\hat{\eta}%
)~,\;\;[\hat{\eta},\hat{\eta}^{\prime }]=i\left. \{\eta ,\eta ^{\prime
}\}_{D(\Phi )}\right| _{\eta =\hat{\eta}}~,\;\;\Phi (\hat{\eta},t)=0\,.
\label{22}
\end{equation}
The latter allows one to introduce a Schr\"{o}dinger picture, where
operators do not depend on time, but the evolution is controlled by the
Schr\"{o}dinger equation with the Hamiltonian $\hat{H}$. We may call the
gauge conditions, which imply the existence of the effective Hamiltonians,
as {\it unitary gauges}. Remember that in the stationary constraint case all
gauge conditions are unitary \cite{GitTy90}. As it is known \cite{GitTy90},
the set of second-class constraints can always be solved explicitly with
respect to part of the variables $\eta _{\ast }=\Psi (\eta ^{\ast }),\hskip%
2mm\eta =(\eta _{\ast },\eta ^{\ast }),$ so that $\eta _{\ast }$ and $\eta
^{\ast }$ are sets of pairs of canonically conjugated variables $\eta _{\ast
}=(q_{\ast },p_{\ast }),\hskip2mm\eta ^{\ast }=(q^{\ast },p^{\ast }).$ We
may call $\eta ^{\ast }$ as independent variables and $\eta _{\ast }$ as
dependent ones. In fact $\eta _{\ast }-\Psi (\eta ^{\ast })=0$ is an
equivalent to $\phi (\eta )=0$ set of second-class constraints. One can
easily demonstrate that it is enough to verify the existence of the
effective Hamiltonian (the validity of relation (\ref{21})) for the
independent variables only. Then the evolution of the dependent variables,
which is controlled by the constraint equations, is also unitary.

Returning to our concrete problem, we remark that in the case under
consideration the Hamiltonian $H$ in the equations (\ref{19}) vanishes
(total Hamiltonian vanishes on the constraint surface). Thus, these
equations take the form 
\begin{equation}
\dot{\eta}=\{\eta ,\epsilon \}_{D(\Phi )}=-\{\eta ,\Phi
_{a}\}C_{ab}\,\partial _{\tau }\Phi _{b},\;\;\Phi _{a}=0\;,  \label{23}
\end{equation}
were $\eta =(x^{\mu },p_{\nu })$, and $C_{ab}\{\Phi _{b},\Phi _{c}\}=\delta
_{ac}$. Calculating the matrices $\{\Phi _{a},\Phi _{b}\}$ and $C_{ab}$ on
the constraint surface, we get $\{\Phi _{a},\Phi _{b}\}={\rm antidiag}%
(-1,1),~C_{ab}=\{\Phi _{b},\Phi _{a}\}$. Let us work now with independent
variables, which are in the case under consideration $\mbox{\boldmath$\y$%
\unboldmath}=(x^{k},p_{k},\zeta )$. It is easy to see that (\ref{23}) imply
the following equations for such variables: 
\begin{equation}
\dot{\mbox{\boldmath$\y$\unboldmath}}=\{\mbox{\boldmath$\y$\unboldmath},%
{\cal H}_{eff}\}\;,\;\;\zeta =\pm 1\;,  \label{24}
\end{equation}
where the effective Hamiltonian $H_{eff}$ reads: 
\begin{equation}
{\cal H}_{eff}=\left[ \zeta qA_{0}(x)+\omega \right] _{x^{0}=\zeta \tau }\;.
\label{25}
\end{equation}
It particular, it follows from (\ref{24}) that $\dot{\zeta}=0$. One can see
that the equations (\ref{24}) are ordinary Hamiltonian equations of motion
without any constraints. Thus, formally, all the problems with
zero-Hamiltonian phenomenon and time dependence of the constraints remain
behind. In fact, we have demonstrated that in the gauge under consideration (%
\ref{15}) the dynamics in the physical sector is unitary and the
corresponding effective Hamiltonian has been constructed explicitly.

\section{First quantization of spinless particle model}

Now, the problem of the canonical operator quantization of the initial gauge
theory is reduced to the quantization of a non-constrained Hamiltonian
theory with the equations (\ref{24}). We assume also the operator $\hat{\zeta%
}$ to have the eigenvalues $\zeta =\pm 1$ by analogy with the classical
theory. The equal time commutation relations for the operators $\hat{X}^{k},%
\hat{P}_{k},\hat{\zeta}$, which correspond to the variables $%
x^{k},p_{k},\zeta $, we define according to their Poisson brackets. Thus,
nonzero commutators are 
\begin{equation}
\lbrack \hat{X}^{k},\hat{P}_{j}]=i\hbar \delta _{j}^{k}\,,\;\,\;\left( \hat{%
\zeta}^{2}=1\right) \;.  \label{a1}
\end{equation}
We are going to present a realization of such an operator algebra in a
Hilbert space and construct there a quantum Hamiltonian $\hat{H}$ according
to the classical expression (\ref{25}).

In the capacity of the above mentioned Hilbert space we select a space $%
{\cal R}$, whose elements $\mbox{\boldmath$\Psi$\unboldmath}\in {\cal R}$
are ${\bf x}$-dependent four-component columns 
\begin{equation}
\mbox{\boldmath$\Psi$\unboldmath}=\left( 
\begin{array}{c}
\Psi _{+1}({\bf x}) \\ 
\Psi _{-1}({\bf x})
\end{array}
\right) \;,\;\;\Psi _{\zeta }({\bf x})=\left( 
\begin{array}{c}
\chi _{\zeta }({\bf x}) \\ 
\varphi _{\zeta }({\bf x})
\end{array}
\right) \;,  \label{a2}
\end{equation}
where $\Psi _{\zeta }({\bf x}),\;\zeta =\pm 1$ are two component columns \
with $\chi $ and $\varphi $ being ${\bf x}$-dependent functions. The inner
product in ${\cal R}$ is defined as follows\footnote{%
Here and in what follows we use standard $\sigma $-matrices, 
\[
\sigma _{1}=\left( 
\begin{array}{cc}
0 & 1 \\ 
1 & 0
\end{array}
\right) ,\;\sigma _{2}=\left( 
\begin{array}{cc}
0 & -i \\ 
i & 0
\end{array}
\right) ,\;\sigma _{3}=\left( 
\begin{array}{cc}
1 & 0 \\ 
0 & -1
\end{array}
\right) ,\; 
\]
}: 
\begin{eqnarray}
&&\left( \mbox{\boldmath$\Psi$\unboldmath},\mbox{\boldmath$\Psi$\unboldmath}%
^{\prime }\right) =\left( \Psi _{+1},\Psi _{+1}^{\prime }\right) +\left(
\Psi _{-1}^{\prime },\Psi _{-1}\right) \;,  \label{a4} \\
&&\left( \Psi ,\Psi ^{\prime }\right) =\int \overline{\Psi }({\bf x})\Psi
^{\prime }({\bf x})d{\bf x}=\int \left[ \chi ^{\ast }({\bf x})\varphi
^{\prime }({\bf x})+\varphi ^{\ast }({\bf x})\chi ^{\prime }({\bf x})\right]
d{\bf x}\;,\;\overline{\Psi }=\Psi ^{+}\sigma _{1}\;.  \label{a5}
\end{eqnarray}
Later on one can see that such a construction of the inner product provides
its form invariance under general coordinate transformations.

We seek all the operators in the block-diagonal form\footnote{%
Here and in what follows we use the following notations 
\[
{\rm bdiag}\left( A,\;B\right) =\left( 
\begin{array}{cc}
A & 0 \\ 
0 & B
\end{array}
\right) \,, 
\]
where $A$ and $B$ are some matrices.}, 
\begin{equation}
\hat{\zeta}={\rm bdiag}\left( I,-I\right) \;,\;\;\hat{X}^{k}=x^{k}{\bf I}%
\;,\;\;\;\hat{P}_{k}=\hat{p}_{k}{\bf I}\;,\;\;\hat{p}_{k}=-i\hbar \partial
_{k}\;,  \label{a6}
\end{equation}
where $I$ and ${\bf I}$ are $2\times 2$ and $4\times 4$ unit matrices
respectively. One can easily see that such defined operators obey the
commutation relations (\ref{a1}) and are Hermitian with respect to the inner
product (\ref{a4}). Evolution of state vectors with the time parameter $\tau 
$ is controlled by the Schr\"{o}dinger equation with a quantum Hamiltonian $%
\hat{H}$. The latter may be constructed as a quantum operator in the Hilbert
space ${\cal R}$ on the base of the correspondence principle starting with
its classical analog, which is ${\cal H}_{eff}$ given by Eq. (\ref{25}).
However, on this way we meet two kind of problems. First of all, one needs
to define the square root in the expression (\ref{8}) on the operator level.
Then one has to solve an ordering problem, it appears due to a
non-commutativity of operators, which have to be situated under the square
root sign in (\ref{8}). Below we are going to discuss both problems. It
seams to be instructive to do that first for a free particle in a flat
space-time, and then in general case (in the presence of both backgrounds,
electromagnetic and gravitational).

For a free particle in a flat space-time $g_{\mu \nu }=\eta _{\mu \nu
},\;A_{\mu }=0\;.$ Then ${\cal H}_{eff}=\omega =\sqrt{m^{2}+(p_{k})^{2}}\;.$
We construct the quantum Hamiltonian as $\hat{H}=\hat{\Omega}$, where $\hat{%
\Omega}$ is an operator related to the classical quantity $\omega $. Such an
operator we define as follows: 
\[
\hat{\Omega}={\rm bdiag}\left( \hat{\omega},\hat{\omega}\right) \;,\;\;\hat{%
\omega}=\left( 
\begin{array}{cc}
0 & m^{2}+(\hat{p}_{k})^{2} \\ 
1 & 0
\end{array}
\right) \;. 
\]
Thus defined operator $\hat{\Omega}$ is Hermitian with respect to the inner
product (\ref{a4}), \ its square $\ \hat{\Omega}^{2}=\left[ m^{2}+(\hat{p}%
_{k})^{2}\right] {\bf I}\;,$ corresponds to the square of the classical
quantity $\omega $, and it is a well defined (in the space ${\cal R}$)
operator function on the basic canonical operators $\hat{p}_{k}$. Thus, the
square root problem is solved here due to the state space doubling (\ref{a2}%
). In the case under consideration we do not meet an ordering problem.

In the general case, when both backgrounds are nontrivial and ${\cal \ H}%
_{eff}$ has the form (\ref{25}), we construct the corresponding quantum
Hamiltonian in the following way: 
\begin{equation}
\hat{H}(\tau )=\hat{\zeta}q\hat{A}_{0}+\hat{\Omega}\;,  \label{a11}
\end{equation}
where the operator $\hat{A}_{0}$ is related to the classical quantity $%
\left. A_{0}\right| _{x^{0}=\zeta \tau }$ and has the following
block-diagonal form $\hat{A}_{0}={\rm bdiag}\left( \left. A_{0}\right|
_{x^{0}=\tau }\,I,\;\left. A_{0}\right| _{x^{0}=-\tau }\,I\,\right) \;,$ and 
$\hat{\Omega}$ is an operator related to the classical quantity $\left.
\omega \right| _{x^{0}=\zeta \tau }$. We define the latter operator as
follows 
\begin{eqnarray}
\hat{\Omega} &=&{\rm bdiag}\left( \left. \hat{\omega}\right| _{x^{0}=\tau
},\;\left. \hat{\omega}\right| _{x^{0}=-\tau }\right) \;,\;\;\hat{\omega}%
=\left( 
\begin{array}{cc}
0 & M \\ 
G & 0
\end{array}
\right) \;,  \label{a13} \\
M &=&-\left[ \hat{p}_{k}+qA_{k}\right] \sqrt{-g}g^{kj}\left[ \hat{p}%
_{j}+qA_{j}\right] +m^{2}\sqrt{-g},\;G=\frac{g_{00}}{\sqrt{-g}}\;.
\label{a14}
\end{eqnarray}
Its square reads $\hat{\Omega}^{2}={\rm bdiag}\left( \left. MG\right|
_{x^{0}=\tau }I,\;\;\left. GM\right| _{x^{0}=-\tau }I\right) ,$ and
corresponds (in the classical limit) to the square of the classical quantity 
$\left. \omega \right| _{x^{0}=\zeta \tau }$. A verification of the latter
statement may be done, for example, on the states with a definite value of $%
\zeta $. Natural symmetric operator ordering in the expression for the
operator $M$ provides the gauge invariance under $U(1)$ transformations of
external electromagnetic field potentials and formal covariance of the
theory under general coordinate transformations as will be seen below. One
can check that the operator $\hat{H}$ is Hermitian with respect to the inner
product (\ref{a4}).

The quantum Hamiltonian (\ref{a11}) may be written in the following
block-diagonal form convenient for the further consideration: 
\begin{equation}
\hat{H}(\tau )={\rm bdiag}\left( \hat{h}(\tau ),\,-\sigma _{3}\hat{h}(-\tau
)\sigma _{3}\right) \,,\;\;\hat{h}(\tau )=\left. \hat{h}(x^{0})\right|
_{x^{0}=\tau },\;\;\hat{h}(x^{0})=qA_{0}I+\hat{\omega}\;.  \label{a17}
\end{equation}
The states of the system under consideration evolute in time $\tau $ in
accordance with the Schr\"{o}dinger equation 
\begin{equation}
i\hbar \partial _{\tau }\mbox{\boldmath$\Psi$\unboldmath}(\tau )=\hat{H}%
(\tau )\mbox{\boldmath$\Psi$\unboldmath}(\tau )\;,  \label{a18}
\end{equation}
where the state vectors $\mbox{\boldmath$\Psi$\unboldmath}$ depend now
parametrically on $\tau $, 
\begin{equation}
\mbox{\boldmath$\Psi$\unboldmath}(\tau )=\left( 
\begin{array}{c}
\Psi _{+1}(\tau ,{\bf x}) \\ 
\Psi _{-1}(\tau ,{\bf x})
\end{array}
\right) ,\;\;\Psi _{\zeta }(\tau ,{\bf x})=\left( 
\begin{array}{c}
\chi _{\zeta }(\tau ,{\bf x}) \\ 
\varphi _{\zeta }(\tau ,{\bf x})
\end{array}
\right) ,\;\;\zeta =\pm 1\;.  \label{a19}
\end{equation}
Taking into account the representation (\ref{a17}), one can see that two
columns $\Psi _{z}(\tau ,{\bf x})$, obey the following equations: 
\begin{equation}
i\hbar \partial _{\tau }\Psi _{+1}(\tau ,{\bf x})=\hat{h}(\tau )\Psi
_{+1}(\tau ,{\bf x})\,,\;i\hbar \partial _{\tau }\Psi _{-1}(\tau ,{\bf x}%
)=-\sigma _{3}\hat{h}(-\tau )\sigma _{3}\Psi _{-1}(\tau ,{\bf x})\;.
\label{a21}
\end{equation}

Let us now demonstrate that the set of equations (\ref{a21}) is equivalent
to two Klein-Gordon equations, one for a scalar field of the charge $q$, and
another one for a scalar field of the charge $-q$. In accordance with our
classical interpretation we may regard $\hat{\zeta}$ as charge sign
operator. Let $\mbox{\boldmath$\Psi$\unboldmath}_{\zeta }$ be states with a
definite charge $\zeta q$, 
\begin{equation}
\hat{\zeta}\mbox{\boldmath$\Psi$\unboldmath}_{\zeta }=\zeta %
\mbox{\boldmath$\Psi$\unboldmath}_{\zeta }\;,\;\;\zeta =\pm 1\;.  \label{a22}
\end{equation}
It is easily to see that states $\mbox{\boldmath$\Psi$\unboldmath}_{+1}$
with the charge $q$ have $\Psi _{-1}=0$. In this case $\tau =x^{0}$, where $%
x^{0}$ is physical time. Then the first equation (\ref{a21}) may be
rewritten as 
\begin{equation}
i\hbar \partial _{0}\Psi _{+1}(x^{0},{\bf x})=\hat{h}(x^{0})\Psi _{+1}(x^{0},%
{\bf x})\,,\;\;\Psi _{+1}(x^{0},{\bf x})=\left( 
\begin{array}{c}
\chi _{+1}(x^{0},{\bf x}) \\ 
\varphi _{+1}(x^{0},{\bf x})
\end{array}
\right) \;.  \label{a23}
\end{equation}
Denoting $\varphi _{+1}(x^{0},{\bf x})=\varphi (x)$ and remembering the
structure (\ref{a13}) of the operator $\hat{\omega}$, we get exactly the
covariant Klein-Gordon equation in curved space-time for the scalar field $%
\varphi (x)$ with the charge $q$, 
\begin{equation}
\left[ \frac{1}{\sqrt{-g}}\left( i\hbar \partial _{\mu }-qA_{\mu }\right) 
\sqrt{-g}g^{\mu \nu }\left( i\hbar \partial _{\nu }-qA_{\nu }\right) -m^{2}%
\right] \varphi (x)=0\;.  \label{a24}
\end{equation}

States $\mbox{\boldmath$\Psi$\unboldmath}_{-1}$ with charge $-q$ have $\Psi
_{+1}=0$. In this case, according to our classical interpretation, $\tau
=-x^{0}$, where $x^{0}$ is physical time. Then, taking into account the
relation 
\begin{equation}
-\left[ \sigma _{3}\hat{h}(x^{0})\sigma _{3}\right] ^{\ast }=\left. \hat{h}%
(x^{0})\right| _{q\rightarrow -q}=\hat{h}^{c}(x^{0})\;.  \label{a25}
\end{equation}
we get from the second equation (\ref{a21}) 
\begin{equation}
i\hbar \partial _{0}\Psi _{-1}^{\ast }(-x^{0},{\bf x})=\hat{h}%
^{c}(x^{0})\Psi _{-1}^{\ast }(-x^{0},{\bf x})\,,\;\;\Psi _{-1}^{\ast
}(-x^{0},{\bf x})=\left( 
\begin{array}{c}
\chi _{-1}^{\ast }(-x^{0},{\bf x}) \\ 
\varphi _{-1}^{\ast }(-x^{0},{\bf x})
\end{array}
\right) \;.  \label{a26}
\end{equation}
Denoting $\varphi _{-1}^{\ast }(-x^{0},{\bf x})=\varphi ^{c}(x)$, one may
rewrite the equation (\ref{a26}) in the form of the covariant Klein-Gordon
equation in curved space-time for the charge conjugated scalar field $%
\varphi ^{c}(x)$ (that which describes particles with the charge $-q$), 
\begin{equation}
\left[ \frac{1}{\sqrt{-g}}\left( i\hbar \partial _{\mu }+qA_{\mu }\right) 
\sqrt{-g}g^{\mu \nu }\left( i\hbar \partial _{\nu }+qA_{\nu }\right) -m^{2}%
\right] \varphi ^{c}(x)=0\;.  \label{a27}
\end{equation}

The inner product (\ref{a4}) between two solutions of the Schr\"{o}dinger
equation (\ref{a18}) with different charges is zero. For two solutions with
charges $q$ it takes the form: 
\begin{eqnarray}
&&\left( \mbox{\boldmath$\Psi$\unboldmath}_{+1},\mbox{\boldmath$\Psi$%
\unboldmath}_{+1}^{\prime }\right) =(\varphi ,\varphi ^{\prime })_{KG}\; 
\nonumber \\
&=&\int \sqrt{-g}g^{00}\left\{ \left[ \left( i\hbar \partial
_{0}-qA_{0}\right) \varphi \right] ^{\ast }\varphi ^{\prime }+\varphi ^{\ast
}\left( i\hbar \partial _{0}-qA_{0}\right) \varphi ^{\prime }\right\} d{\bf x%
}\,,  \label{a28}
\end{eqnarray}
it is expressed via Klein-Gordon scalar product on the $x^{0}={\rm const}$
hyperplane for the case of the charge $q$ , see (\ref{b11}). For two
solutions with charges $-q$ the inner product (\ref{a4}) reads: 
\begin{eqnarray}
&&\left( \mbox{\boldmath$\Psi$\unboldmath}_{-1},\mbox{\boldmath$\Psi$%
\unboldmath}_{-1}^{\prime }\right) =(\varphi ^{c},\varphi ^{c^{\prime
}})_{KG}^{c}\;  \nonumber \\
&=&\int \sqrt{-g}g^{00}\left\{ \left[ \left( i\hbar \partial
_{0}+qA_{0}\right) \varphi ^{c}\right] ^{\ast }\varphi ^{c^{\prime
}}+\varphi ^{c\ast }\left( i\hbar \partial _{0}+qA_{0}\right) \varphi
^{c^{\prime }}\right\} d{\bf x}\,,  \label{a29}
\end{eqnarray}
it is expressed via Klein-Gordon scalar product for the case of the charge $%
-q$, which is denoted above by an upper index $c$.

Each block-diagonal operator $\hat{T}$, which acts in ${\cal R}$, induces
operators acting on the fields $\varphi (x)$ and $\varphi ^{c}(x)$. In
particular, for the operators $\hat{{\cal P}}_{k}=\hat{\zeta}\hat{P}_{k}$ of
the physical momenta in a flat space-time (see classical interpretation), we
get$\hat{\;{\cal P}}_{k}\mbox{\boldmath$\Psi$\unboldmath}\;\rightarrow \hat{p%
}_{k}\;\varphi (x)$ and $\hat{p}_{k}\;\varphi ^{c}(x)\;,$ as one can expect,
since the form of the momentum operators do not depend on the field charge
sign.

The above demonstration, together with the previous classical analysis (see
Sect.II) has confirmed ones again that $x^{0}$ may be treated as physical
time. Thus, it is natural to reformulate the evolution in the quantum
mechanics constructed in terms of this physical time. At the same time we
pass to a different representation of state vectors, taking into account the
physical meaning of the components $\Psi _{\pm 1}$, which follows from the
equation (\ref{a23}) and (\ref{a26}). Namely, we will describe quantum
mechanical states by means of four columns 
\begin{eqnarray}
&&\mbox{\boldmath$\Psi$\unboldmath}(x^{0})=\left( 
\begin{array}{c}
\Psi (x) \\ 
\Psi _{c}(x)
\end{array}
\right) \,,\;\;\Psi (x)=\Psi _{+1}(x^{0},{\bf x}),\;\Psi ^{c}(x)=\Psi
_{-1}^{\ast }(-x^{0},{\bf x})\,,  \nonumber \\
&&\Psi (x)=\left( 
\begin{array}{c}
\chi (x) \\ 
\varphi (x)
\end{array}
\right) \,,\;\;\Psi ^{c}(x)=\left( 
\begin{array}{c}
\chi ^{c}(x) \\ 
\varphi ^{c}(x)
\end{array}
\right) \;.  \label{a31}
\end{eqnarray}
As was said above it is, in fact, a transition to a new representation. Such
a representation we may call conditionally $x^{0}$-representation, in
contrast with the representation (\ref{a2}) or (\ref{a19}), which will be
called $\tau $-representation. The inner product of two states $%
\mbox{\boldmath$\Psi$\unboldmath}(x^{0})$ and $\mbox{\boldmath$\Psi$%
\unboldmath}^{\prime }(x^{0})$ in $x^{0}$ representation takes the form 
\begin{equation}
\left( \mbox{\boldmath$\Psi$\unboldmath},\mbox{\boldmath$\Psi$\unboldmath}%
^{\prime }\right) =\left( \Psi ,\Psi ^{\prime }\right) +\left( \Psi ^{c},{%
\Psi ^{c}}^{\prime }\right) \;,  \label{a32}
\end{equation}
where the product $\left( \Psi ,\Psi ^{\prime }\right) $ is still given by
the equation (\ref{a5}).

One may find expressions for the basic operators in $x^{0}$-representation
under consideration. The operators $\hat{\zeta}$ and $\hat{X}^{k}$ defined
by the expressions (\ref{a6}) retain their form, whereas the expressions for
the Hamiltonian and the momenta change. The former one has the form 
\begin{equation}
\hat{H}(x^{0})={\rm bdiag}\left( \hat{h}(x^{0})\,,\;\ \hat{h}%
^{c}(x^{0})\right) \;,  \label{a33}
\end{equation}
where $\hat{h}(x^{0})$ is the corresponding Hamiltonian from (\ref{a17}) and 
$\hat{h}^{c}(x^{0})$ is given by Eq. (\ref{a25}). The operator of the
physical momentum $\hat{{\cal P}}_{i}=\zeta \hat{P}_{i}$ has the form $\hat{%
{\cal P}}_{i}=$\ \ $\hat{p_{k}}{\bf I}$\ in $x^{0}$ representation. That
confirms the interpretation of the physical momentum derived from the
classical consideration in the previous Section. The time evolution of state
vectors in $x^{0}$-representation follows from the equations (\ref{a23}) and
(\ref{a26}) 
\begin{equation}
i\hbar \partial _{0}\mbox{\boldmath$\Psi$\unboldmath}(x^{0})=\hat{H}(x^{0})%
\mbox{\boldmath$\Psi$\unboldmath}(x^{0})\;.  \label{a35}
\end{equation}

\section{First quantized theory and one-particle sector of quantized scalar
field}

Below we will give an interpretation of the quantum mechanics constructed,
comparing it with a dynamics of a one-particle sector of QFT of complex
scalar field. To this end we are going first to demonstrate that the
one-particle sector of the QFT (in cases when it may be consistently
defined, see Appendix) may be formulated as a relativistic quantum mechanics
without infinite number of negative energy levels and negative norms of
state vectors. Then we will show that it may be identified (under certain
suppositions) with the quantum mechanics, which was constructed by us in the
previous Section in course of the first quantization of the corresponding
classical action. Doing that, we may, at the same time, give a more exact
interpretation of the quantum mechanics. Below we use widely notions,
results, and notations presented in the Appendix.

To begin with one ought to remember that the one-particle sector of QFT (as
well as any sector with a definite particle number) may be defined in an
unique way for all time instants only in external backgrounds, which do not
create particles from the vacuum \cite
{BirDa82,GreMuR85,GriMaM88,Fulli89,FraGiS91}. Nonsingular time independent
external backgrounds give an important example of the above backgrounds, see
Appendix. That is why we are going first to present a discussion for such
kind of backgrounds to simplify the consideration. A generalization to
arbitrary backgrounds, in which the vacuum remains stable, may be done in
the similar manner.

Let us reduce the total Fock space ${\cal R}^{FT}$ of the QFT to a subspace
of vectors which obey the condition $\hat{N}|\mbox{\boldmath$\Psi$%
\unboldmath}>=|\mbox{\boldmath$\Psi$\unboldmath}>,$ where $\hat{N}$ \ is the
operator of number of particles (\ref{b57}). Namely, we select the subspace $%
{\cal R}^{1}={\cal R}_{10}^{FT}\oplus {\cal R}_{01}^{FT}$ of normalized
vectors having the form: 
\begin{equation}
|\mbox{\boldmath$\Psi$\unboldmath}>=\sum_{n}\left[ f_{n}a_{n}^{+}+\lambda
_{n}b_{n}^{+}\right] |0>\;,  \label{c3}
\end{equation}
where $f_{n},\;\lambda _{n}$ are arbitrary coefficients, $\sum_{n}\left[
|f_{n}|^{2}+|\lambda _{n}|^{2}\right] <\infty $. We are going to call ${\cal %
R}^{1}$ one-particle sector of QFT. All state vectors from the one-particle
sector (as well as any vector from the Fock space) have positive norms.

The vectors $|n,\zeta >$ form a complete basis in ${\cal R}^{1}$, 
\begin{equation}
|n,\zeta >=\left\{ 
\begin{array}{c}
a_{n}^{+}|0>,\;\zeta =+1\;, \\ 
b_{n}^{+}|0>,\;\zeta =-1.
\end{array}
\right.  \label{c4}
\end{equation}
The spectrum of the Hamiltonian $\hat{H}_{R}^{FT}$ (see (\ref{b52})) in the
space ${\cal R}^{1}$ reproduces exactly one-particle energy spectrum of
particles and antiparticles without infinite number of negative energy
levels, 
\begin{equation}
\hat{H}_{R}^{FT}|n,+1>=\epsilon _{+,n}|n,+1>,\;\hat{H}_{R}^{FT}|n,-1>=%
\epsilon _{+,n}^{c}|n,-1>\;.  \label{c5}
\end{equation}

The dynamics of the one-particle sector may be formulate as a relativistic
quantum mechanics under certain suppositions. To demonstrate that, we pass
first to a coordinate representation for state vectors of the QFT, which is
an analog of common in nonrelativistic quantum mechanics coordinate
representation. Consider the decompositions 
\begin{eqnarray}
&&\hat{\psi}(x)=\hat{\psi}_{(-)}(x)+\hat{\psi}_{(+)}(x),\;\;\hat{\psi}%
^{c}(x)=-\left( \hat{\psi}^{+}\sigma _{3}\right) ^{T}=\hat{\psi}%
_{(-)}^{c}(x)+\hat{\psi}_{(+)}^{c}(x)\;,  \nonumber \\
&&\hat{\psi}_{(-)}(x)=\sum_{n}a_{n}\psi _{+,n}(x),\;\hat{\psi}%
_{(+)}(x)=\sum_{\alpha }b_{\alpha }^{+}\psi _{-,\alpha }(x)\;,  \nonumber \\
&&\hat{\psi}_{(-)}^{c}(x)=-\sum_{\alpha }b_{\alpha }\sigma _{3}\psi
_{-,\alpha }^{\ast }(x)=\sum_{\alpha }b_{\alpha }\psi _{+,\alpha }^{c}(x)\;,
\nonumber \\
&&\hat{\psi}_{(+)}^{c}(x)=-\sum_{n}a_{n}^{+}\sigma _{3}\psi _{+,n}^{\ast
}(x)=\sum_{n}a_{n}^{+}\psi _{-,n}^{c}(x)\;,  \label{c6}
\end{eqnarray}
where $\hat{\psi}^{c}(x)$ is charge conjugated Heisenberg operator of the
field $\hat{\psi}(x)$. By means of such defined operators we may construct a
ket basis in the Fock space ${\cal R}^{FT}$, 
\begin{equation}
<0|\hat{\psi}_{(-)}({\bf x}_{1})\ldots \hat{\psi}_{(-)}({\bf x}_{A})\hat{\psi%
}_{(-)}^{c}({\bf y}_{1})\ldots \hat{\psi}_{(-)}^{c}({\bf y}%
_{B})\,,\;\;A,B=0,1,...\;,  \label{c7}
\end{equation}
where 
\[
\hat{\psi}_{(-)}({\bf x})=\left. \hat{\psi}_{(-)}(x)\right| _{x^{0}=0},\;\;%
\hat{\psi}_{(-)}^{c}({\bf y})=\left. \hat{\psi}_{(-)}^{c}(y)\right|
_{y^{0}=0}\;. 
\]
Then a time-dependent state vector $|\mbox{\boldmath$\Psi$\unboldmath}%
(x^{0})>$ from the Fock space (in the coordinate representation) is given by
a set of its components 
\begin{eqnarray}
&&\Psi _{AB}(x_{1}\ldots x_{A},y_{1}\ldots y_{B})=<0|\hat{\psi}_{(-)}({\bf x}%
_{1})\ldots \hat{\psi}_{(-)}({\bf x}_{A})\hat{\psi}_{(-)}^{c}({\bf y}%
_{1})\ldots \hat{\psi}_{(-)}^{c}({\bf y}_{B})|\mbox{\boldmath$\Psi$%
\unboldmath}(x^{0})>  \nonumber \\
&=&<0|\hat{\psi}_{(-)}(x_{1})\ldots \hat{\psi}_{(-)}(x_{A})\hat{\psi}%
_{(-)}^{c}(y_{1})\ldots \hat{\psi}_{(-)}^{c}(y_{B})|\mbox{\boldmath$\Psi$%
\unboldmath}(0)>\;,  \nonumber \\
&&x_{1}^{0}=\ldots =x_{A}^{0}=y_{1}^{0}=\ldots =y_{B}^{0}=x^{0}\;.
\label{c8}
\end{eqnarray}

Let us consider a time-dependent state $|\mbox{\boldmath$\Psi$\unboldmath}%
(x^{0})>$ from the subspace ${\cal R}^{1}$. It has only nonzero components $%
\Psi _{10}(x)$ and $\Psi _{01}(x)$, 
\begin{eqnarray}
&&\Psi _{10}(x)=<0|\hat{\psi}_{(-)}(x)|\mbox{\boldmath$\Psi$\unboldmath}%
(0)>=<0|\hat{\psi}(x)|\mbox{\boldmath$\Psi$\unboldmath}(0)>\;,  \nonumber \\
&&\Psi _{01}(x)=<0|\hat{\psi}_{(-)}^{c}(x)|\mbox{\boldmath$\Psi$\unboldmath}%
(0)>=<0|\hat{\psi}^{c}(x)|\mbox{\boldmath$\Psi$\unboldmath}(0)>\;.
\label{c9}
\end{eqnarray}
Thus, one may describe states from ${\cal R}^{1}$ in the coordinate
representation by four columns 
\begin{eqnarray}
&&\mbox{\boldmath$\Psi$\unboldmath}(x^{0})=\left( 
\begin{array}{c}
\Psi (x) \\ 
\Psi ^{c}(x)
\end{array}
\right) ,\;\Psi (x)=\Psi _{10}(x),\;\Psi ^{c}(x)=\Psi _{01}(x),  \nonumber \\
&&\Psi (x)=\left( 
\begin{array}{c}
\chi (x) \\ 
\varphi (x)
\end{array}
\right) ,\;\Psi ^{c}(x)=\left( 
\begin{array}{c}
\chi ^{c}(x) \\ 
\varphi ^{c}(x)
\end{array}
\right) \;.  \label{c10}
\end{eqnarray}

The QFT inner product $<\mbox{\boldmath$\Psi$\unboldmath}|%
\mbox{\boldmath$\Psi$\unboldmath}^{\prime }>$ of two states from ${\cal R}%
^{1}$ may be written via their representatives in the coordinate
representation. To this end one may use the following expression for the
projection operator to one-particle sector 
\begin{equation}
\int \left[ \hat{\overline{\psi }}({\bf x})|0><0|\hat{\psi}({\bf x})+\hat{%
\overline{\psi ^{c}}}({\bf x})|0><0|\hat{\psi}^{c}({\bf x})\right] d{\bf x}%
=I\;.  \label{c11}
\end{equation}
It follows from the relation (\ref{b20}) and properties of the solutions $%
\psi _{\zeta ,n}(x)$. Then the inner product $\left( \mbox{\boldmath$\Psi$%
\unboldmath},\mbox{\boldmath$\Psi$\unboldmath}^{\prime }\right) $ of two
states $\mbox{\boldmath$\Psi$\unboldmath}(x^{0})$ and $\mbox{\boldmath$\Psi$%
\unboldmath}^{\prime }(x^{0})$ from the one-particle sector in the
coordinate representation may be written as 
\begin{eqnarray}
&&\left( \mbox{\boldmath$\Psi$\unboldmath},\mbox{\boldmath$\Psi$\unboldmath}%
^{\prime }\right) =\left( \Psi ,\Psi ^{\prime }\right) +\left( \Psi ^{c},{%
\Psi ^{c}}^{\prime }\right) ,\;\left( \left( \mbox{\boldmath$\Psi$%
\unboldmath},\mbox{\boldmath$\Psi$\unboldmath}^{\prime }\right) =<%
\mbox{\boldmath$\Psi$\unboldmath}|\mbox{\boldmath$\Psi$\unboldmath}^{\prime
}>\right) ,  \nonumber \\
&&\left( \Psi ,\Psi ^{\prime }\right) =\int \overline{\Psi }(x)\Psi (x)d{\bf %
x}=\int \left[ \chi ^{\ast }(x)\varphi ^{\prime }(x)+\varphi ^{\ast }(x)\chi
^{\prime }(x)\right] d{\bf x}\;.  \label{c12}
\end{eqnarray}

One may find expressions for the basic operators in the coordinate
representation in the one-particle sector: 
\begin{equation}
\hat{H}_{R}^{FT}\rightarrow \hat{H}={\rm bdiag}\left( \hat{h}\,,\;\hat{h}%
^{c}\right) \;,  \label{c13}
\end{equation}
where $\hat{h}$ and $\hat{h}^{c}$ are defined by Eq. (\ref{a17}) and (\ref
{a25}), so that (\ref{c13}) is, in fact, the quantum mechanical Hamiltonian (%
\ref{a33}) in the case under consideration (in time-independent
backgrounds); $\hat{P}_{k}^{FT}\rightarrow \hat{P}_{k}={\rm bdiag}\left(
P_{k},\;P_{k}^{c}\right) \;,$ where the operator $P_{k}$ is defined by the
expression (\ref{b1}), $P_{k}=i\hbar \partial _{k}-qA_{k},$ and $%
P_{k}^{c}=\left. P_{k}\right| _{q\rightarrow -q}=-P_{k}^{\ast }=i\hbar
\partial _{k}+qA_{k}$; $\hat{Q}^{FT}\rightarrow \hat{Q}=q\hat{\zeta}\;,$ \
where $\hat{\zeta}$ is the operator from (\ref{a6}).

An analog of the equations (\ref{c5}) in the coordinate representation has
the form 
\begin{eqnarray}
&&\hat{H}\mbox{\boldmath$\Psi$\unboldmath}_{+,n}=\epsilon _{+,n}%
\mbox{\boldmath$\Psi$\unboldmath}_{+,n},\;\;\mbox{\boldmath$\Psi$\unboldmath}%
_{+,n}=\left( 
\begin{array}{c}
\psi _{+,n}({\bf x}) \\ 
0
\end{array}
\right) ,  \nonumber \\
&&\hat{H}\mbox{\boldmath$\Psi$\unboldmath}_{+,\alpha }^{c}=\epsilon
_{+,\alpha }^{c}\mbox{\boldmath$\Psi$\unboldmath}_{+,\alpha }^{c},\;\;%
\mbox{\boldmath$\Psi$\unboldmath}_{+,\alpha }^{c}=\left( 
\begin{array}{c}
0 \\ 
\psi _{+,\alpha }^{c}({\bf x})
\end{array}
\right) ,  \nonumber \\
\; &&\;\left( \mbox{\boldmath$\Psi$\unboldmath}_{+,n},\mbox{\boldmath$\Psi$%
\unboldmath}_{+,m}\right) =\delta _{nm},\left( \mbox{\boldmath$\Psi$%
\unboldmath}_{+,\alpha }^{c},\mbox{\boldmath$\Psi$\unboldmath}_{+,\beta
}^{c}\right) =\delta _{\alpha \beta },\left( \mbox{\boldmath$\Psi$%
\unboldmath}_{+,n},\mbox{\boldmath$\Psi$\unboldmath}_{+,\alpha }^{c}\right)
=0,\;\;  \label{c16}
\end{eqnarray}
see (\ref{b33}), (\ref{b36}), and (\ref{b39}). The set $\mbox{\boldmath$%
\Psi$\unboldmath}_{+,n},\;\mbox{\boldmath$\Psi$\unboldmath}_{+,\alpha }^{c}$
forms a complete basis in ${\cal R}^{1}$ in the coordinate representation.

The time evolution of state vectors from the one-particle sector in the
coordinate representation may be found using the equations (\ref{b49}).
Thus, one may write 
\begin{equation}
i\hbar \partial _{0}\Psi (x)=\hat{h}\Psi (x)\,,\;\;i\hbar \partial _{0}\Psi
^{c}(x)=\hat{h}^{c}\Psi ^{c}(x)\;,  \label{c17}
\end{equation}
or using the notations introduced above 
\begin{equation}
i\hbar \partial _{0}\mbox{\boldmath$\Psi$\unboldmath}(x^{0})=\hat{H}%
\mbox{\boldmath$\Psi$\unboldmath}(x^{0})\;.  \label{c18}
\end{equation}

According to superselection rules \cite{Weinb95} physical states are only
those which are eigenvectors for the charge operator (\ref{b52}). Thus,
among the vectors (\ref{c3}) only those, which obey the condition 
\begin{equation}
\hat{Q}^{FT}|\mbox{\boldmath$\Psi$\unboldmath}>=\zeta q|\mbox{\boldmath$%
\Psi$\unboldmath}>,\;\;\zeta =\pm 1\;,  \label{c20}
\end{equation}
are physical. This condition defines a physical subspace ${\cal R}_{ph}^{1}$
from the one-particle sector. It is easy to see that ${\cal R}_{ph}^{1}=%
{\cal R}_{10}^{FT}\cup {\cal R}_{01}^{FT}\;.$ \ Vectors from the physical
subspace of the one-particle sector have the form: 
\begin{equation}
|\mbox{\boldmath$\Psi$\unboldmath}>=\left(
\sum_{n}f_{n}a_{n}^{+}|0>,\;\sum_{n}\lambda _{n}b_{n}^{+}|0>\right) \;,
\label{c22}
\end{equation}
where $f_{n},\;\lambda _{n}$ are arbitrary coefficients, $%
\sum_{n}|f_{n}|^{2}<\infty ,\;|\lambda _{n}|^{2}<\infty $. Since the charge
operator has the block-diagonal form in the one-particle sector in the
coordinate representation (see above), the condition (\ref{c20}) reads 
\begin{equation}
\hat{\zeta}\mbox{\boldmath$\Psi$\unboldmath}_{\zeta }=\zeta %
\mbox{\boldmath$\Psi$\unboldmath}_{\zeta },\;\;\zeta =\pm 1\;.  \label{c23}
\end{equation}
Thus, the physical subspace of the one-particle sector consists of the
vectors $\mbox{\boldmath$\Psi$\unboldmath}_{\zeta },\;\;\zeta =\pm 1$ only.
Due to the structure of the operator $\hat{\zeta}$, the states $%
\mbox{\boldmath$\Psi$\unboldmath}_{+1}$ contain only the upper half of
components, whereas, ones $\mbox{\boldmath$\Psi$\unboldmath}_{-1}$ contain
only the lower half of components, 
\begin{equation}
\mbox{\boldmath$\Psi$\unboldmath}_{+1}=\left( 
\begin{array}{c}
\Psi ({\bf x}) \\ 
0
\end{array}
\right) ,\;\;\mbox{\boldmath$\Psi$\unboldmath}_{-1}=\left( 
\begin{array}{c}
0 \\ 
\Psi ^{c}({\bf x})
\end{array}
\right) \;.  \label{c24}
\end{equation}
One may see that the complete set $\mbox{\boldmath$\Psi$\unboldmath}_{+,n}$
and $\mbox{\boldmath$\Psi$\unboldmath}_{+,\alpha }^{c}$ from (\ref{c16})
consists only of physical vectors.

The continuity equation, which follows from (\ref{c18}), has the form 
\begin{eqnarray}
&&\partial _{0}\rho +{\rm div}{\bf j}=0,\;\rho =\overline{\Psi }\Psi +%
\overline{\Psi ^{c}}\Psi ^{c}\;,  \nonumber \\
&&j^{i}=\frac{1}{2}g^{ik}\sqrt{-g}\left\{ \left[ \overline{\Psi }P_{k}+%
\overline{(P_{k}\Psi )}\right] (\sigma _{1}+i\sigma _{2})\Psi +\left[ 
\overline{\Psi ^{c}}P_{k}^{c}+\overline{(P_{k}^{c}\Psi ^{c})}\right] (\sigma
_{1}+i\sigma _{2})\Psi ^{c}\right\} \;.  \label{c19}
\end{eqnarray}
Let us denote via $\rho _{\zeta }(x)$ the quantity (\ref{c19}) constructed
from the physical states $\mbox{\boldmath$\Psi$\unboldmath}_{\zeta }$. This
quantity may not be interpreted as a probability density, since it is not
positively defined in general case. That is a reflection of a well-known
fact that one cannot construct one-particle localized states in relativistic
quantum theory. However, due to the Eq. (\ref{c12}) we get $\int \rho
_{\zeta }d{\bf x}=\left( \mbox{\boldmath$\Psi$\unboldmath}_{\zeta },%
\mbox{\boldmath$\Psi$\unboldmath}_{\zeta }\right) =1$ \ for any normalized
physical states. Moreover, the overlap $\left( \mbox{\boldmath$\Psi$%
\unboldmath}_{\zeta },\mbox{\boldmath$\Psi$\unboldmath}_{\zeta }^{\prime
}\right) $ may be treated as a probability amplitude, supporting usual
quantum mechanical interpretation.

Summarizing all what was said above, we may conclude that, in fact, the QFT
dynamics in the physical subspace of the one-particle sector in the
coordinate representation is formulated as a consistent relativistic quantum
mechanics. It does not meet well-known difficulties usual for standard
formulations of the relativistic quantum mechanics of spinless particles 
\cite{FesVi58,Schwe61,Grein97} such as negative norms and infinite number of
negative energy levels.

Now we may return to the interpretation of the results of the first
quantization presented in the previous Section. In the time independent
nonsingular backgrounds under consideration, we may see that under certain
restrictions our quantum mechanics coincides literally with the dynamics of
the QFT in the physical subspace of the one-particle sector. These
restrictions are related only to an appropriate definition of the Hilbert
space of the quantum mechanics. Indeed, all other constructions in the
quantum mechanics and in the one-particle sector of the QFT in the
coordinate representation coincide. The space ${\cal R}$, in which the
commutation relations (\ref{a1}) were realized (the space of the vectors of
the form (\ref{a2})), is too wide, in particular, it contains negative norm
vectors. We have first to restrict it to a subspace, which is equivalent to $%
{\cal R}^{1}$ and after that to the physical subspace ${\cal R}_{ph}^{1}$.
Thus, we may get a complete equivalence between the both theories. To do the
first step we consider the eigenvalue problem for the Hamiltonian (\ref{a33}%
) in the space ${\cal R}$. Its spectrum is wider than one (\ref{c16}) in the
space ${\cal R}^{1}$, 
\begin{eqnarray}
\hat{H}\mbox{\boldmath$\Psi$\unboldmath}_{\varkappa ,n} &=&\epsilon
_{\varkappa ,n}\mbox{\boldmath$\Psi$\unboldmath}_{\varkappa ,n},\;%
\mbox{\boldmath$\Psi$\unboldmath}_{\varkappa ,n}=\left( 
\begin{array}{c}
\psi _{\varkappa ,n}({\bf x}) \\ 
0
\end{array}
\right) ,\;\left( \mbox{\boldmath$\Psi$\unboldmath}_{\varkappa ,n},%
\mbox{\boldmath$\Psi$\unboldmath}_{\varkappa ^{\prime },m}\right) =\varkappa
\delta _{\varkappa \varkappa ^{\prime }}\delta _{nm\,},  \nonumber \\
\hat{H}\mbox{\boldmath$\Psi$\unboldmath}_{\varkappa ,\alpha }^{c}
&=&\epsilon _{\varkappa ,\alpha }^{c}\mbox{\boldmath$\Psi$\unboldmath}%
_{\varkappa ,\alpha }^{c},\;\mbox{\boldmath$\Psi$\unboldmath}_{\varkappa
,\alpha }^{c}=\left( 
\begin{array}{c}
0 \\ 
\psi _{\varkappa ,\alpha }^{c}({\bf x})
\end{array}
\right) ,\;\left( \mbox{\boldmath$\Psi$\unboldmath}_{\varkappa ,\alpha }^{c},%
\mbox{\boldmath$\Psi$\unboldmath}_{\varkappa ^{\prime },\beta }^{c}\right)
=\varkappa \delta _{\varkappa \varkappa ^{\prime }}\delta _{\alpha \beta }\,,
\nonumber \\
&&\left( \mbox{\boldmath$\Psi$\unboldmath}_{\varkappa ,n},%
\mbox{\boldmath$\Psi$\unboldmath}_{\varkappa ^{\prime },\alpha }^{c}\right)
=0\;,\varkappa ,\varkappa ^{\prime }=\pm \;,  \label{c26}
\end{eqnarray}
see (\ref{b33}), (\ref{b36}), and (\ref{b39}). To get the same spectrum as
in QFT, we need to eliminate all the vectors $\mbox{\boldmath$\Psi$%
\unboldmath}_{-,n}$ and $\mbox{\boldmath$\Psi$\unboldmath}_{-,\alpha }^{c}$
from the consideration. Thus, we may define the analog of the space ${\cal R}%
^{1}$ as a linear envelop of the vectors $\mbox{\boldmath$\Psi$\unboldmath}%
_{+,n}$ and $\mbox{\boldmath$\Psi$\unboldmath}_{+,\alpha }^{c}$ only. This
space does not contain negative norm vectors and the operator $\hat{\Omega}$
from (\ref{a13}) is positively defined in perfect accordance with the
positivity of the classical quantity $\omega $. The spectrum of the
Hamiltonian in such defined space coincides with one of the Hamiltonian of
the QFT in the one-particle sector. Reducing ${\cal R}^{1}$ to ${\cal R}%
_{ph}^{1}$, we get literal coincidence between both theories.

One ought to mention a well-know in the relativistic quantum mechanics
problem of position operator construction (see \cite
{FolWo50,FesVi58,Schwe61,GroSu72} and references there). In all the works
where they started with a given K-G or Dirac equation as a Schr\"{o}dinger
ones, the construction of such an operator was a heuristic task. The form of
the operator had to be guessed to obey some physical demands, by analogy
with the nonrelativistic case. In particular, an invariance of the
one-particle sector with a given sign of energy under the action of the
position operator was expected. Besides, mean values of the operator had to
have necessary transformation properties under the coordinate
transformations and the correspondence principle had to hold. Realizing
these and some other demands they met serious difficulties. At present, from
the position of a more deep understanding of the quantum field theory, it is
clear that it is impossible to construct localized one-particle states. That
means that the position operator with the above mentioned properties does
not exist. In the frame of our consideration, which starts with a given
classical theory, the coordinate $x$ becomes an operator $\hat{X}$ in course
of the quantization . Thus, the correspondence principle holds
automatically. We do not demand from the operator $\hat{X}$ \ literally
similar properties as in non-relativistic quantum mechanics. In particular,
the one-particle sector is not invariant under the action of such an
operator. The operator has no eigenvectors in this sector. To construct such
eigenvectors one has to include into the consideration many-particle states.
In the present work we do not exceed the limits of the one-particle
consideration. However, a generalization to many-particle theory may be done
one the base of the constructed one-particle sector and the existence of
eigenvectors of the operator $\hat{X}$ may be demonstrated. That will be
presented in our next article. As to the momentum operator, similar problem
appears only in nonuniform external backgrounds, and has to be understood
similarly.

The above comparison of the first quantized theory with the dynamics of the
one-particle sector of QFT was done for non-singular and time-independent
external backgrounds. It may be easily extended to any time-dependent
background, which do not create particles from vacuum.

Thus, we see that the first quantization of a classical action leads to a
relativistic quantum mechanics which is consistent to the same extent as
quantum field theory in the one-particle sector. Such a quantum mechanics
describes spinless charged particles of both signs, and reproduces correctly
their energy spectra, which is placed on the upper half-plane of the Fig.1
(see App.).

One may think that the reduction of the space ${\cal R}$ of the quantum
mechanics to the space ${\cal R}^{1}$ is necessary only in the first
quantization, thus an equivalence between the first and the second
quantization is not complete. That may be interpreted as a weak point in the
presented scheme of the first quantization. However, it is a wrong
impression, the same procedure is present in the second quantization. Below
we are going to remember how it happens in the case under consideration of
scalar field. Indeed, instead to write the decomposition (A34) one could
write two possible decompositions 
\[
\hat{\psi}(x)=\sum_{\varkappa n}a_{\varkappa ,n}\psi _{\varkappa
,n}(x)\,,\;\ \hat{\psi}^{c}(x)=\sum_{\varkappa n}b_{\varkappa ,n}\psi
_{\varkappa ,n}^{c}(x)\;, 
\]
since both sets $\psi _{\varkappa ,n}$ and $\psi _{\varkappa ,n}^{c}$ are
complete. Then it follows from the commutation relations (A32) that

\[
\lbrack a_{\varkappa ,n},a_{\varkappa ^{\prime },m}^{+}]=[b_{\varkappa
,n},b_{\varkappa ^{\prime },m}^{+}]=\varkappa \delta _{\varkappa \varkappa
^{\prime }}\delta _{nm},\;[a_{\varkappa ,n},a_{\varkappa ^{\prime
},m}]=[b_{\varkappa ,n},b_{\varkappa ^{\prime },m}]=0\,. 
\]
Two sets of operators $a,a^{+}$ and $b,b^{+}$ are not independent, they are
related as follows: $a_{+,n}=b_{-,n}^{+}\,,\;a_{+,n}^{+}=b_{-,n}\,,%
\;a_{-,n}=b_{+,n}^{+}\,,\;a_{-,n}^{+}=b_{+,n}\;.$ Interpreting the operators
without cross as annihilation ones, we may define four vacuum vectors

\[
|0\rangle =|a_{+}\rangle \otimes |b_{+}\rangle \,,\;|0\rangle
_{1}=|a_{+}\rangle \otimes |a_{-}\rangle \,,\;|0\rangle _{2}=|b_{+}\rangle
\otimes |b_{-}\rangle \,,\;|0\rangle _{3}=|a_{-}\rangle \otimes
|b_{-}\rangle \,, 
\]
where $a_{+,n}|a_{+}>=0,\;a_{-,n}|a_{-}>=0,\;b_{+,n}|b_{+}>=0,%
\;b_{-,n}|b_{-}>=0\,,$ and the following one-particle excited states:

\[
1)\;a_{+,n}^{+}|0\rangle \,,\;b_{+,n}^{+}|0\rangle
\,;\;2)\;\;a_{+,n}^{+}|0\rangle _{1}\,,\;a_{-,n}^{+}|0\rangle
_{1}\,;\;3)\;b_{+,n}^{+}|0\rangle _{2}\,,\;b_{-,n}^{+}|0\rangle
_{2}\,;\;4)\;\;a_{-,n}^{+}|0\rangle _{3}\,,\;b_{-,n}^{+}|0\rangle _{3}\,. 
\]
The non-renormalized quantum Hamiltonian, which may be constructed from the
classical expression (A14), reads $\hat{H}^{FT}=\sum_{\varkappa ,n}\varkappa
\epsilon _{\varkappa ,n}a_{\varkappa ,n}^{+}a_{\varkappa ,n}$ . Now one may
see that only the one-particle states from the group 1 form the physical
subspace. All other states from the groups 2,3,4 have to be eliminated,
since they or contain negative energy levels, negative norms, or do not
reproduce complete spectrum of particles and antiparticles. Working in such
defined physical subspace we may deal only with the operators$%
\;a_{+,n}^{+},a_{+,n},b_{+,n}^{+},b_{+,n}$ and denote them simply as $%
a_{n}^{+},a_{n},b_{n}^{+},b_{n}.$ Then all usual results of second quantized
theory may be reproduced.

Finally, to complete the consideration let us examine the Dirac quantization
of the theory in question. In this case we do not need to impose any gauge
condition to the first-class constraint (\ref{9}). We assume as before the
operator $\hat{\zeta}$ to have the eigenvalues $\zeta =\pm 1$ by analogy
with the classical theory. The equal time commutation relations for the
operators $\hat{X}^{\mu },\hat{P}_{\mu },\hat{\zeta}$, which correspond to
the variables $x^{\mu },p_{\mu },\zeta $, we define according to their
Poisson brackets, due \ to the absence of second-class constraints. Thus,
now we get for nonzero commutators 
\begin{equation}
\ [\hat{X}^{\mu },\hat{P}_{\nu }]=i\hbar \delta _{\nu }^{\mu }\;,\;\;\left( 
\hat{\zeta}^{2}=1\;\right) .  \label{c27}
\end{equation}
Besides, we have to keep in mind the necessity to construct an operator
realization for the first class-constraint (\ref{9}), which contains a
square root. Taking all that into account, we select as a state space one
whose elements $\mbox{\boldmath$\Psi$\unboldmath}$ are $x$-dependent
four-component columns 
\begin{equation}
\mbox{\boldmath$\Psi$\unboldmath}=\left( 
\begin{array}{c}
\Psi _{+1}(x) \\ 
\Psi _{-1}(x)
\end{array}
\right) \;,\;\;\Psi _{\zeta }(x)=\left( 
\begin{array}{c}
\chi _{\zeta }(x) \\ 
\varphi _{\zeta }(x)
\end{array}
\right) \;,  \label{c28}
\end{equation}
where $\Psi _{\zeta }(x),\;\zeta =\pm 1$ are two component columns, \ with $%
\chi $ and $\varphi $ being $x$-dependent functions. We seek all the
operators in the block-diagonal form, 
\begin{equation}
\hat{\zeta}={\rm bdiag}\left( I,\;-I\right) \;,\;\;\hat{X}^{\mu }=x^{\mu }%
{\bf I}\;,\;\;\;\hat{P}_{\mu }=\hat{p}_{\mu }{\bf I}\;,\;\;\hat{p}_{\mu
}=-i\hbar \partial _{\mu }\;.  \label{c30}
\end{equation}
where $I$ and ${\bf I}$ are $2\times 2$ and $4\times 4$ unit matrices
respectively. The operator $\hat{\Phi}_{1}$ which corresponds to the
first-class constraint (\ref{9}), is selected as $\hat{\Phi}_{1}=\hat{P}%
_{0}+q\hat{A}_{0}+\hat{\zeta}\hat{\Omega}$ . The operators $\hat{A}_{0}$ and 
$\hat{\Omega}\ $ related to the classical quantities $A_{0}$ and $\ \omega $
have the following forms $\;\hat{A}_{0}=A_{0}{\bf I},\;\;\hat{\Omega}={\rm %
bdiag}\left( \hat{\omega},\hat{\omega}\right) \,,$ where $\hat{\omega}$ is
defined by Eq. (\ref{a13}). Similar to the canonical quantization case, one
may verify that the square $\hat{\Omega}^{2}$ corresponds (in the classical
limit) to the square of the classical quantity $\omega $. The state vectors (%
\ref{c28}) do not depend on ''time'' $\tau $ since the Hamiltonian vanishes
on the constraints surface. The physical state vectors have to obey the
equation $\hat{\Phi}_{1}\mbox{\boldmath$\Psi$\unboldmath}=0\,.$ Thus, we
arrive to the equations

\begin{equation}
i\hbar \partial _{0}\Psi _{+1}(x)=\left( qA_{0}+\hat{\omega}\right) \Psi
_{+1}(x)\;,\;\;i\hbar \partial _{0}\Psi _{-1}(x)=\left( qA_{0}-\hat{\omega}%
\right) \Psi _{-1}(x)\;.  \label{c32}
\end{equation}
Taking into account the realization of all the operators, definitions (\ref
{a17}), (\ref{a25}), and denoting $\Psi _{+1}(x)=\psi (x),\;-\sigma _{3}\Psi
_{-1}^{\ast }(x)=\psi ^{c}(x),$ we get two Klein-Gordon equations (in the
Hamiltonian form)

\begin{equation}
i\partial _{0}\psi =\hat{h}(x^{0})\psi \;,\;\;i\partial _{0}\psi ^{c}=\hat{h}%
^{c}(x^{0})\psi ^{c}\,,  \label{c33}
\end{equation}
one for particle, and one for the antiparticle, see (\ref{b17}) and (\ref
{b24}). Unfortunately, the Dirac method of the quantization gives no more
information how to proceed further with the consistent quantum theory
construction, and moreover contains principal contradictions, see discussion
in the Introduction. However, we may conclude that at least one of the main
feature of the quantum theory, its charge conjugation invariance, remains
also in the frame of the Dirac quantization.

\section{Spinning particle case}

We would like to demonstrate here that a consistent quantization, similar to
that for spinless particle, applied to an action of spinning particle,
allows one to construct a consistent relativistic quantum mechanics, which
is equivalent to one-particle sector of quantized spinor field. For
simplicity we restrict ourselves here only by one external electromagnetic
background, considering the problem in the flat space-time.

An action of spin one half relativistic particle (spinning particle), with
spinning degrees of freedom describing by anticommuting (Grassmann-odd)
variables, was first proposed by Berezin and Marinov \cite{BM} and just
after that discussed and studied in detail in papers \cite
{Casa,BCL,BDZVH,BVH,BSSW}. It may be written in the following form (in the
flat space-time) 
\begin{equation}
S=\int_{0}^{1}Ld\tau \,,\;\ L=-\frac{\left( \dot{x}^{\mu }-i\xi ^{\mu }\chi
\right) ^{2}}{2e}-q\dot{x}^{\mu }A_{\mu }+iqeF_{\mu \nu }\xi ^{\mu }\xi
^{\nu }-im\xi ^{4}\chi -\frac{e}{2}m^{2}-i\xi _{n}\dot{\xi}^{n}\;,
\label{s1}
\end{equation}
where $x^{\mu },~e$ are even and $\xi ^{n},\;\chi $ are odd variables,
dependent on a parameter $\tau \in \lbrack 0,1]$, which plays a role of time
in this theory, $\mu ={\overline{0,3}};~n=(\mu ,4)=\overline{0,4};~\eta
_{\mu \nu }={\rm diag}(1,-1,-1,-1);~\eta _{mn}={\rm diag}(1,-1,-1,-1,-1).$
Spinning degrees of freedom are described by odd variables $\xi ^{n}$; even $%
e$ and odd $\chi $ play an auxiliary role to make the action
reparametrization and super gauge-invariant as well as to make it possible
consider both cases $m\neq 0$ and $m=0$ on the same foot.

The are two types of gauge transformations under which the action (\ref{s1})
is invariant: reparametrizations $\delta x^{\mu }={\dot{x}}^{\mu
}\varepsilon \;,\;\delta e=\frac{d}{d\tau }\left( e\varepsilon \right)
\;,\;\delta \xi ^{n}=\dot{\xi}^{n}\varepsilon \;,\;\delta {\chi }=\frac{d}{%
d\tau }\left( \chi \varepsilon \right) \;,$ and supertransformations $\delta
x^{\mu }=i\xi ^{\mu }\epsilon \;,~~\delta e=i\chi \epsilon \;,\;\;\delta
\chi =\dot{\epsilon}\,,\;\delta \xi ^{\mu }=\frac{1}{2e}\left( \dot{x}^{\mu
}-i\xi ^{\mu }\chi \right) \epsilon \;,~~\delta \xi ^{4}=\frac{m}{2}\epsilon
\;,$ where $\varepsilon (\tau )\,$and$\ \epsilon (\tau )$ are $\tau $%
-dependent gauge parameter, the first one is even and the second one is odd.

Going over to Hamiltonian formulation, we introduce the canonical momenta: 
\begin{equation}
p_{\mu }=\frac{\partial L}{\partial \dot{x}^{\mu }}=-\frac{\dot{x}_{\mu
}-i\xi _{\mu }\chi }{e}-qA_{\mu }\,,\;P_{e}=\frac{\partial L}{\partial \dot{e%
}}=0\,,\;P_{\chi }=\frac{\partial _{r}L}{\partial \dot{\chi}}=0\,,\;\pi _{n}=%
\frac{\partial _{r}L}{\partial \dot{\xi}^{n}}=-i\xi _{n}\,.  \label{s5}
\end{equation}
It follows from (\ref{s5}) that there exist primary constraints ${\phi }%
^{(1)}=0$, 
\begin{equation}
\phi _{1}^{(1)}=P_{\chi }\,,\;\phi _{2}^{(1)}=P_{e}\,,\;\phi
_{3,n}^{(1)}=\pi _{n}+i\xi _{n}\;.  \label{s6}
\end{equation}
We construct the total Hamiltonian $H^{(1)}=H+\lambda _{a}\phi _{a}^{(1)},$
according to standard procedure \cite{GitTy90,Dirac64,HenTe92}, 
\begin{equation}
H=-\frac{e}{2}\left[ \left( p+qA\right) ^{2}-m^{2}+2iqF_{\mu \nu }\xi ^{\mu
}\xi ^{\nu }\right] +i\left[ \left( p_{\mu }+qA_{\mu }\right) \xi ^{\mu
}+m\xi ^{4}\right] \chi \;.  \label{s7}
\end{equation}
From the conditions of the conservation of the primary constraints $\phi
^{(1)}$ in the time $\tau ,\;\dot{\phi}^{(1)}=\left\{ \phi
^{(1)},H^{(1)}\right\} =0$, we find secondary constraints $\phi ^{(2)}=0$, 
\begin{equation}
\phi _{1}^{(2)}=(p_{\mu }+qA_{\mu })\xi ^{\mu }+m\xi ^{4}\,,\;\;\phi
_{2}^{(2)}=(p+qA)^{2}-m^{2}+2iqF_{\mu \nu }\xi ^{\mu }\xi ^{\nu }\;,
\label{s8}
\end{equation}
and determine $\lambda $, which correspond to the primary constraint $\phi
_{3n}^{(1)}$. Thus, the Hamiltonian $H$ appears to be proportional to the
constraints, as one could expect in the case of a reparametrization
invariant theory, $H=-\frac{e}{2}\phi _{2}^{(2)}+i\phi _{1}^{(2)}\chi \;.$
No more secondary constraints arise from the Dirac procedure, and the
Lagrange multipliers, corresponding to the primary constraints $\phi
_{1}^{(1)},\;\phi _{2}^{(1)},$ remain undetermined.

One can go over from the initial set of constraints $\phi ^{(1)},\phi ^{(2)}$
to the equivalent one $\phi ^{(1)},T$, where $\ $ 
\begin{eqnarray}
&&T_{1}=\left( p_{\mu }+qA_{\mu }\right) \left( \pi ^{\mu }-i\xi ^{\mu
}\right) +m\left( \pi ^{4}-i\xi ^{4}\right) \;,  \label{s9} \\
&&T_{2}=p_{0}+qA_{0}+\zeta r\;,\;\;r=\sqrt{m^{2}+\left( p_{k}+qA_{k}\right)
^{2}+2qF_{\mu \nu }\xi ^{\mu }\pi ^{\nu }}\,.  \label{s10}
\end{eqnarray}
The new set of constraints can be explicitly divided in a set of the
first-class constraints, which is $\phi _{1,2}^{(1)},\;T$, and in a set of
second-class constraints, which is $\phi _{3,n}^{(1)}$, 
\begin{equation}
\left\{ \phi _{a}^{(1)},\phi ^{(1)}\right\} =\left\{ \phi
_{a}^{(1)},T\right\} =\left. \left\{ T,\phi _{3,n}^{(1)}\right\} \right|
_{\phi =T=0}=\left. \left\{ T,T\right\} \right| _{\phi =T=0}=0\;,\;\;a=1,2\,.
\label{s11}
\end{equation}
The constraint (\ref{s10}) is equivalent to one $\phi _{2}^{(2)},\;$ $\phi
_{2}^{(2)}=-2\zeta rT_{2}+\left( T_{2}\right) ^{2}$ . Remember, that $\zeta
=-{\rm sign}\left[ p_{0}+qA_{0}(x)\right] \;.$ Thus, the constraint (\ref
{s10}) is a analog of the linearized primary constraint (\ref{9}) in the
scalar particle case.

We are going to impose supplementary gauge conditions to all the first-class
constraints. First we impose two gauge conditions $\phi ^{G}=0$, 
\begin{equation}
\phi _{1}^{G}=\pi ^{0}-i\xi ^{0}+\zeta \left( \pi ^{4}-i\xi ^{4}\right)
,\;\;\phi _{2}^{G}=x^{0}-\zeta \tau \;\;.  \label{s12}
\end{equation}
A motivation for the gauge condition $\phi _{2}^{G}$ is the same as in
scalar particle case (Sect. II) . As to the gauge condition $\phi _{1}^{G}$,
it is chosen to be a contrpart to one of the first-class constraint $T$ ,
and to provide a simple structure of the final complete set of second-class
constraints, see below. It differs from similar gauge condition, which was
used in \cite{GriGr95}, by a combination of constraints. From the
consistency condition $\dot{\phi}^{G}=0$ we find two additional constraints

\begin{eqnarray}
\phi _{3}^{G} &=&\chi -\frac{i\zeta qF_{k0}\left( \pi ^{k}-i\xi ^{k}\right) 
}{\tilde{\omega}_{0}(\tilde{\omega}_{0}+m)}=0\;,  \label{s13} \\
\phi _{4}^{G} &=&e-\frac{1}{\tilde{\omega}}\left[ 1-\frac{iqF_{k0}\left( \pi
^{k}-i\xi ^{k}\right) \left( \pi ^{0}-i\xi ^{0}\right) }{2\tilde{\omega}_{0}(%
\tilde{\omega}_{0}+m)}\right] =0,  \label{s14}
\end{eqnarray}
where 
\begin{equation}
\tilde{\omega}_{0}=\sqrt{m^{2}+\left( p_{k}+qA_{k}\right) ^{2}+2qF_{kl}\xi
^{k}\pi ^{l}}\,,\;\tilde{\omega}=\sqrt{\tilde{\omega}_{0}^{2}+\frac{2\zeta
qF_{k0}}{\tilde{\omega}_{0}+m}\left( p_{l}+qA_{l}\right) \left( \xi ^{k}\pi
^{l}+\pi ^{k}\xi ^{l}\right) }\;.  \label{s15}
\end{equation}
Then, the conditions of consistency for the constraints of $\phi _{3}^{G}$
and $\phi _{4}^{G}$ lead to the determination of the Lagrange multipliers
for the primary constraints $\phi _{1}^{(1)}$ and $\phi _{2}^{(1)}$. The
complete set of constraints $\left( \phi ^{(1)},T,\phi ^{G}\right) $ is
already a second-class one.

Below we are going to present an equivalent to $\left( \phi ^{(1)},T,\phi
^{G}\right) $ set of second-class constraints $\Phi _{a},\;a=1,2,...,13,$
which has a simple quasi-diagonal matrix $\left\{ \Phi _{a},\Phi
_{b}\right\} $. The first five constraints of this set have the form 
\begin{equation}
\Phi _{1}=p_{0}+qA_{0}+\zeta \tilde{\omega}\,,\;\Phi _{2}=\phi
_{2}^{G}\,,\;\Phi _{3}=\phi _{3,1}^{(1)}\,,\;\Phi _{4}=\phi
_{3,2}^{(1)}\,,\;\;\,\Phi _{5}=\phi _{3,3\;}^{(1)}\;.  \label{s16}
\end{equation}
Four of them are exactly old constraints, and the first one is a linear
combination of the old constraints. Namely, $\Phi
_{1}=t_{1}T_{1}+t_{2}T_{2}+f\phi _{1}^{G}+f_{nm}\phi _{3,m}^{(1)}\phi
_{3,n}^{\left( 1\right) }$ \ , where the coefficient functions are 
\begin{eqnarray*}
t_{1} &=&\frac{-i\zeta qF_{k0}\left( \pi ^{k}-i\xi ^{k}\right) }{(m+\tilde{%
\omega}_{0})(p_{0}+qA_{0}-\zeta \tilde{\omega}\,)}\;,\;\;f=\frac{%
imqF_{k0}(\pi ^{k}-i\xi ^{k})}{(m+\tilde{\omega}_{0})(p_{0}+qA_{0}-\zeta 
\tilde{\omega}\,)}\,,\; \\
t_{2} &=&\frac{1}{p_{0}+qA_{0}-\zeta \tilde{\omega}\,\ }\left[
p_{0}+qA_{0}-\zeta r\;+\frac{i\zeta qF_{k0}\left( \pi ^{k}-i\xi ^{k}\right)
\left( \pi ^{0}-i\xi ^{0}\right) }{(m+\tilde{\omega}_{0})}\right] \,, \\
f_{k,0} &=&\frac{iqF_{k0}}{p_{0}+qA_{0}-\zeta \tilde{\omega}\,}\left[ 1-%
\frac{iqF_{l0}\left( \pi ^{l}-i\xi ^{l}\right) \left( \pi ^{0}-i\xi
^{0}\right) }{2(m+\tilde{\omega}_{0})\tilde{\omega}_{0}}\right] \,,\;f_{k,l}=%
\frac{i\zeta qF_{k0}\left( p_{l}+qA_{l}\right) }{(m+\tilde{\omega}%
_{0})(p_{0}+qA_{0}-\zeta \tilde{\omega}\,)}\,.\;
\end{eqnarray*}
The rest constraints are orthogonal (in sense of the Poisson brackets)\ to
the latter five and form four orthogonal to each other pairs. The first pair
is $\Phi _{6},\,\Phi _{7}\,,\;$where$\;\Phi _{6}=-\frac{i}{2}%
T_{1}+bT_{2}+c\phi _{2}^{G}+r_{k}\phi _{3,k}^{(1)}\phi _{3,0}^{(1)},\;\;\Phi
_{7}=\phi _{1}^{G}\;,$ and 
\[
b=\frac{i\left\{ \phi _{2}^{G},T_{1}\right\} }{2\left\{ \phi
_{2}^{G},T_{2}\right\} }\,,\;\;c=-\frac{\left\{ -\frac{i}{2}%
T_{1}+bT_{2}+r_{k}\phi _{3,k}^{(1)}\phi _{3,0}^{(1)},\Phi _{1}\right\} }{%
\left\{ \phi _{2}^{G},\Phi _{1}\right\} }\,,\;\;r_{k}=\frac{\left( \pi
^{0}-i\xi ^{0}\right) \zeta qF_{k0}}{4\tilde{\omega}_{0}}\;. 
\]
The second pair is $\Phi _{8},\,\Phi _{9}\,,$ where $\Phi _{8}=\phi
_{3}^{G}+d\phi _{2}^{G}+v\phi _{1}^{G}+u\;\phi _{2}^{\left( 1\right)
},\;\;\;\ \Phi _{9}=\phi _{1}^{(1)}\,,$ and 
\[
d=-\frac{\left\{ \phi _{3}^{G},\Phi _{1}\right\} }{\left\{ \phi
_{2}^{G},\Phi _{1}\right\} }\,,\;\;v=-\frac{\left\{ \phi _{3}^{G},\Phi
_{6}\right\} }{\left\{ \Phi _{7},\Phi _{6}\right\} }\,,\;\;u=-\frac{\left\{
\phi _{3}^{G}+v\phi _{1}^{\left( 1\right) },\phi _{4}^{G}\right\} }{\left\{
\phi _{2}^{(1)},\phi _{4}^{G}\right\} }\;. 
\]
The third pair is $\Phi _{10},\,\Phi _{11}\,,$ where $\Phi _{10}=\phi
_{4}^{G}+w\phi _{2}^{G}+z\Phi _{7}+s\Phi _{6}\,,\;\;\;\Phi _{11}=\phi
_{2}^{(1)}\,,$ and 
\[
w=-\frac{\left\{ \phi _{4}^{G},\Phi _{1}\right\} }{\left\{ \phi
_{2}^{G},\Phi _{1}\right\} }\,,\;\;z=-\frac{\left\{ \phi _{4}^{G},\Phi
_{6}\right\} }{\left\{ \Phi _{7},\Phi _{6}\right\} }\,,\;\;s=-\frac{\left\{
\phi _{4}^{G},\Phi _{7}\right\} }{\left\{ \Phi _{6},\Phi _{7}\right\} }\;. 
\]
The last pair is $\Phi _{12},\,\Phi _{13}\,,$ where $\Phi _{12}=\phi
_{3,0}^{\left( 1\right) },\;\;\;\ \Phi _{13}=\phi _{3,4}^{(1)}\,\,.$ \ All
nonzero \ Poisson brackets \ between the new constraints are listed below 
\begin{eqnarray}
\left\{ \Phi _{2},\Phi _{1}\right\} &=&-\left\{ \Phi _{1},\Phi _{2}\right\}
=1\,,\;\;\left\{ \Phi _{3},\Phi _{3}\right\} =\left\{ \Phi _{4},\Phi
_{4}\right\} =\left\{ \Phi _{5},\Phi _{5}\right\} =-2i\;,\;  \nonumber \\
\left\{ \Phi _{6},\Phi _{7}\right\} &=&\left\{ \Phi _{7},\Phi _{6}\right\}
=\zeta (\tilde{\omega}_{0}+m)\,,\;\;\left\{ \Phi _{8},\Phi _{9}\right\}
=\left\{ \Phi _{9},\Phi _{8}\right\} =1\;,  \nonumber \\
\left\{ \Phi _{10},\Phi _{11}\right\} &=&-\left\{ \Phi _{11},\Phi
_{10}\right\} =1\,,\;\;\left\{ \Phi _{12},\Phi _{12}\right\} =-\left\{ \Phi
_{13},\Phi _{13}\right\} =2i\,\;.  \label{s21}
\end{eqnarray}
Now we are in position to analyze the equations of motion in the case under
consideration. They have the form (\ref{19}), in which one has to put $H=0$, 
\begin{equation}
\dot{\eta}=\left\{ \eta ,\varepsilon \right\} _{D(\Phi )}\,,\;\;\Phi =0\;,
\label{s22}
\end{equation}
where $\eta $ stands for the set of all the variables of the theory, and the
Dirac brackets are considered in the extended phase space (see Sect. II). We
are going to demonstrate that an effective Hamiltonian exists in this case.
To this end let us divide the complete set of constraints $\Phi $ into two
subsets of constraints, $U$ and $V$ , $\Phi =\left( U,V\right) $, where $%
U=\left( \Phi _{a}\right) \,,\;\ a=1,...,5\,,\;\;V=\left( \Phi _{b}\right)
\,,\;\;b=6,...,13\;.$ \ It is easy to see that both $U$ and $V$ are sets of
second-class constraints. In this case the Dirac brackets with respect to
the constraints $\Phi $ may be calculated successively (see \cite{GitTy90},
p.276), 
\begin{equation}
\left\{ {\cal F},{\cal G}\right\} _{D(\Phi )}=\left\{ {\cal F},{\cal G}%
\right\} _{D(U)}-\left\{ {\cal F},V_{b}\right\} _{D(U)}C^{bd}\left\{ V_{d},%
{\cal G}\right\} _{D(U)}\,,  \label{s24}
\end{equation}
where $C^{^{bd}}\left\{ V_{d},V_{c}\right\} _{D(U)}=\delta _{c}^{b}$ and $%
{\cal F},\;{\cal G}$ are some functions on phase variables. Consider only
variables ${\mbox{\boldmath$\y$\unboldmath}}=\left( x^{k},p_{k},\zeta ,\xi
^{k},\pi _{k}\right) $ . All other variables may be expressed via these
variables, or eliminate from the consideration by means of constraints.
Applying the formula (\ref{s24}), and taking into account the specific
structure of the constraints $V$, we may write equations of motion for the
variables ${\mbox{\boldmath$\y$\unboldmath}}$ in the following simple form 
\begin{equation}
\dot{\mbox{\boldmath$\y$\unboldmath}}=\left\{ {\mbox{\boldmath$\y$%
\unboldmath}}{\Bbb ,}\varepsilon \right\} _{D(U)\,,\;\;}U=0\;.  \label{s25}
\end{equation}
Now let us divide the complete set of constraints $U$ into two subsets of
constraints, $u$ and $v,$ $U=\left( u,v\right) $, where $u=\left( \Phi
_{a}\right) \,,\;\ a=3,4,5\,,\;\;v=\left( \Phi _{b}\right) \,,\;\;b=1,2\;.$
It is easy to see that both $u$ and $v$ are sets of second-class
constraints. Now we may again calculate the Dirac brackets from Eq. (\ref
{s25}) successively. Here a simplification comes from the fact that $\left\{ 
{\mbox{\boldmath$\y$\unboldmath}}{\Bbb ,}\varepsilon \right\} _{D(u)\,\ }=0$%
. Thus we get 
\begin{equation}
\left\{ {\mbox{\boldmath$\y$\unboldmath}}{\Bbb ,}\varepsilon \right\}
_{D(U)}=-\left\{ {\mbox{\boldmath$\y$\unboldmath},}v_{a}\right\}
_{D(u)}c^{ab}\left\{ v_{b},\varepsilon \right\} _{D(u)}\,,\;\
\;c^{ab}\left\{ v_{b},v_{d}\right\} _{D(u)}=\delta _{d\,\,.}^{a}  \label{s27}
\end{equation}
The matrix $c$ may be easily calculated: $c^{11}=c^{22}=0,\;c^{12}=-c^{21}=1%
\,.$ Then the above equation may be written as 
\begin{equation}
\left\{ {\mbox{\boldmath$\y$\unboldmath}}{\Bbb ,}\varepsilon \right\}
_{D(U)}=\left\{ {\mbox{\boldmath$\y$\unboldmath},}\zeta \Phi _{1}\right\}
_{D(u)}\,.  \label{s28}
\end{equation}
The term $\zeta \Phi _{1}$ under the Dirac bracket sign in (\ref{s28}) may
be transformed in the following way: First we may eliminate the momentum $%
p_{0}$ from $\Phi _{1}$ (there is no $x^{0}$ in ${\mbox{\boldmath$\y$%
\unboldmath}}$ ), then substitute $x^{0}$ by $\zeta \tau $ according to the
constraint $\Phi _{2}=0$ (there is no $p_{0}$ in ${\mbox{\boldmath$\y$%
\unboldmath}}$ and in $u$), and finally to express all the momenta $\pi _{k}$
by $-i\xi _{k}$ according to the constraints $u=0$ (that may be done since
the Dirac brackets are just taken with respect of the constraints $u$).
Thus, finally we may write the equations of motion for the variables ${%
\mbox{\boldmath$\y$\unboldmath}}$ in the following form 
\begin{equation}
\dot{\mbox{\boldmath$\y$\unboldmath}}=\{{\mbox{\boldmath$\y$\unboldmath}},%
{\cal H}_{eff}\}_{D(u)}\,,\;\;u_{k}=\phi _{3,k}^{(1)}=0\,\,,\;\ k=1,2,3\;\;.
\label{s29}
\end{equation}
where the effective Hamiltonian $H_{eff}$ reads: 
\begin{eqnarray}
{\cal H}_{eff} &=&\left[ \zeta qA_{0}+\omega \right] _{x^{0}=\zeta \tau
}\;,\;\;\;\omega =\left. \tilde{\omega}\right| _{\pi _{k}=-i\xi _{k}}=\sqrt{%
\omega _{0}^{2}+\rho }\;,  \nonumber \\
\omega _{0} &=&\sqrt{m^{2}+\left( p_{k}+qA_{k}\right) ^{2}-2iqF_{kl}\xi
^{k}\xi ^{l}}\;,\;\;\ \rho =\frac{-4i\zeta qF_{k0}}{\omega _{0}+m}%
(p_{l}+qA_{l})\xi ^{k}\xi ^{l}\;.  \label{s30}
\end{eqnarray}
The nonzero Dirac brackets between the independent variables ${%
\mbox{\boldmath$\y$\unboldmath}}$ have the form 
\[
\left\{ x^{k},p_{l}\right\} _{D(u)}=\left\{ x^{k},p_{l}\right\} =\delta
_{l\,,}^{k}\;\;\left\{ \xi ^{k},\xi ^{l}\right\} _{D(u)}=\frac{i}{2}\eta
^{kl}\;. 
\]
Then the equal time commutation relations for the operators $\hat{X}^{k},%
\hat{P}_{k},\hat{\zeta}$,$\hat{\Xi}^{k},$ which correspond to the variables $%
x^{k},p_{k},\zeta $,$\xi ^{k},$ we define according to their Dirac brackets.
The nonzero commutators (anticommutators) are 
\begin{equation}
\ [\hat{X}^{k},\hat{P}_{j}]=i\hbar \delta _{j}^{k}\,,\;\;[\hat{\Xi}^{k},\hat{%
\Xi}^{l}]_{+}=-\frac{\hbar }{2}\eta ^{kl}\;.  \label{s31}
\end{equation}
We assume as before $\hat{\zeta}^{2}=1$ , and select a state space ${\cal R}$
whose elements $\mbox{\boldmath$\Psi$\unboldmath}\in {\cal R}$ are ${\bf x}$%
-dependent eight-component columns 
\begin{equation}
\mbox{\boldmath$\Psi$\unboldmath}=\left( 
\begin{array}{c}
\Psi _{+1}({\bf x}) \\ 
\Psi _{-1}({\bf x})
\end{array}
\right) \;,  \label{s32}
\end{equation}
where $\Psi _{\zeta }({\bf x}),\;\zeta =\pm 1$ are four component columns.
The inner product in ${\cal R}$ is defined as follows: 
\begin{equation}
\left( \mbox{\boldmath$\Psi$\unboldmath},\mbox{\boldmath$\Psi$\unboldmath}%
^{\prime }\right) =\left( \Psi _{+1},\Psi _{+1}^{\prime }\right) +\left(
\Psi _{-1}^{\prime },\Psi _{-1}\right) \,,\;\left( \Psi ,\Psi ^{\prime
}\right) =\int \Psi ^{\dagger }({\bf x})\Psi ({\bf x})d{\bf x}\;.
\label{s33}
\end{equation}
Later on one can see that such a construction of the inner product provides
its form invariance under Lorenz transformations. We seek all the operators
in the block-diagonal form, in particular, the operators $\hat{\zeta}$ and $%
\hat{\Xi}^{k}$ we chose as: 
\begin{equation}
\hat{\zeta}={\rm bdiag}\left( I,\;-I\right) \;,\;\;\hat{\Xi}^{k}={\rm bdiag}%
\left( \hat{\xi}^{k},\;\hat{\xi}^{k}\right) \,,\;\left[ \hat{\xi}^{k},\hat{%
\xi}^{l}\right] _{+}=-\frac{\hbar }{2}\eta ^{kl}\;,  \label{s34}
\end{equation}
where $I$ is $4\times 4$ unit matrix, and $\hat{\xi}^{k}$ are some $4\times
4 $ matrices, which obey the above equal time commutation relation. Thus, we
may realize the operators $\hat{\xi}^{k}$ by means of $\gamma $ -matrices, 
\begin{equation}
\hat{\xi}^{k}=\frac{i}{2}\hbar ^{1/2}\gamma ^{k}\;,\;\;\left[ \gamma
^{k},\gamma ^{l}\right] _{+}=2\eta ^{kl}\;.  \label{s35}
\end{equation}
The canonical operators $\hat{X}^{k}$ and $\hat{P}_{k}$ we define as totally
diagonal 
\begin{equation}
\hat{X}^{k}=x^{k}{\bf I}\;,\;\;\;\hat{P}_{k}=\hat{p}_{k}{\bf I}\;,\;\;\hat{p}%
_{k}=-i\hbar \partial _{k}\;,  \label{s36}
\end{equation}
where ${\bf I}$ are $8\times 8$ unit matrix. One can easily see that such
defined operators really obey the commutation relations (\ref{s31}) and are
Hermitian with respect to the inner product (\ref{s33}). Evolution of state
vectors with the time parameter $\tau $ is controlled by a Schr\"{o}dinger
equation with a quantum Hamiltonian $\hat{H}$. The latter may be constructed
as a quantum operator in the Hilbert space ${\cal R}$ on the base of the
correspondence principle starting with its classical analog, which is ${\cal %
H}_{eff}$ given by Eq. (\ref{s30}). There exist infinite number of possible
operators which have the same classical image. That corresponds to the
well-known ambiguity of the quantization in general case. We construct the
corresponding quantum Hamiltonian in the following way: 
\begin{equation}
\hat{H}(\tau )=\hat{\zeta}q\hat{A}_{0}+\hat{\Omega}\;.  \label{s37}
\end{equation}
The operator $\hat{A}_{0}$ has the following diagonal form $\hat{A}_{0}={\rm %
bdiag}\left( \left. A_{0}\right| _{x^{0}=\tau }\,I,\;\;\left. A_{0}\right|
_{x^{0}=-\tau }\,I\right) \;.$ \ We define the operator $\hat{\Omega}$ as
follows 
\begin{equation}
\hat{\Omega}={\rm bdiag}\left( \left. \hat{\omega}_{0}\right| _{x^{0}=\tau
\,,\;}-\left. \hat{\omega}_{0}\right| _{x^{0}=-\tau }\right) \;,\;\;\;\hat{%
\omega}_{0}=\gamma ^{0}\left[ m+\gamma ^{k}\left( \hat{p}_{k}+qA_{k}\right) 
\right] \;,\;  \label{s39}
\end{equation}
where $\gamma ^{0}$ is one of the Dirac matrix, $\left( \gamma ^{0}\right)
^{2}=1,\;\;\left[ \gamma ^{0},\gamma ^{k}\right] _{+}=0\;.$ The first term
in the expression (\ref{s37}) is a natural quantum image of the classical
quantity $\left. \zeta qA_{0}\right| _{x^{0}=\zeta \tau }$ . Below we are
going to adduce some arguments demonstrating that the second term $\hat{%
\Omega}$ may be considered as a quantum image of the classical quantity $%
\left. \omega \right| _{_{x^{0}=\zeta \tau }}$ . In fact, we have to justify
the following symbolic relation 
\begin{equation}
\lim_{classical}\hat{\Omega}=\left. \omega \right| _{_{x^{0}=\zeta \tau }}\;.
\label{s40}
\end{equation}
To be more rigorous one has to work with operator symbols. However, we
remain here in terms of operators, hoping that our manipulations have a
clear sense and do not need to be confirmed on the symbol language. First,
we may replace the operator $\hat{\Omega}$ under the sign of the limit by
another one $\hat{\Omega}^{\prime }=\hat{\Omega}+\hat{\Delta}$ , where 
\[
\hat{\Delta}={\rm bdiag}\left( \left. \gamma ^{0}\hat{\xi}^{k}\hat{\lambda}%
_{k}\right| _{x^{0}=\tau },\;\left. \gamma ^{0}\hat{\xi}^{k}\hat{\lambda}%
_{k}\right| _{x^{0}=-\tau }\right) \,,\;\;\hat{\lambda}_{k}=-\hbar ^{-1/2}m%
\left[ \left[ qF_{l0}\hat{\xi}^{l},\frac{1}{\hat{\omega}_{0}^{2}-m^{2}}%
\right] _{+},\hat{\xi}_{k}\right] \,.\; 
\]
Indeed, one may see that the classical limit of the operator $\hat{\Delta}$
is zero. A justification may be the following: The leading in $\hbar $
contribution to the operator $\hat{\Delta}$ results from the terms, which
contain $\left( \hat{\xi}^{k}\right) ^{2}$ . Such operators have classical
limit zero. That is related, for example, to the fact that due to the
realization (\ref{s35}) they are proportional to $\hbar $. On the other
side, we may remember \ that in the classical limit such terms turn out to
be proportional to $\left( \xi ^{k}\right) ^{2}$, which is zero due to
Grassmann nature of $\xi $ . Both considerations are consistent. As to the
operator $\hat{\Omega}^{\prime }$ , we may consider its square and see that
its classical limit corresponds to the square of the classical quantity $%
\left. \omega \right| _{_{x^{0}=\zeta \tau }}\;.$ It would be enough to
prove the relation (\ref{s40}). The concrete details look as follows 
\begin{eqnarray*}
&&\left( \hat{\Omega}^{\prime }\right) ^{2}={\rm bdiag}\left( \left. \hat{%
\omega}^{2}\right| _{x^{0}=\tau ,\zeta =1},\left. \hat{\omega}^{2}\right|
_{x^{0}=-\tau ,\zeta =-1}\right) ,\;\hat{\omega}^{2}=\hat{\omega}_{0}^{2}+%
\hat{\rho}_{1}+\hat{\rho}_{2}\,,\;\hat{\omega}_{0}^{2}=\left[ m^{2}+(\hat{p}%
_{k}+qA_{k})^{2}\right] I \\
&&-iqF_{jl}[\hat{\xi}^{j},\hat{\xi}^{l}]\;,\;\hat{\rho}_{1}=\frac{1}{2i}%
\left[ \left[ \zeta qF_{k0}\hat{\xi}^{k},\frac{1}{\hat{\omega}_{0}^{2}-m^{2}}%
\right] _{+},\left[ \left( \hat{p}_{j}+qA_{j}\right) \hat{\xi}^{j},\hat{%
\omega}_{0}-m\right] _{+}\right] \;,\; \\
&&\hat{\rho}_{2}=\frac{m}{2i}\left[ \left[ \left[ \zeta qF_{k0}\hat{\xi}^{k},%
\frac{1}{\hat{\omega}_{0}^{2}-m^{2}}\right] _{+},\hat{\xi}^{j}\right]
_{+},\left( \hat{p}_{j}+qA_{j}\right) \right] -\frac{\hbar \hat{\lambda}%
_{k}^{2}}{4}-\frac{[\hat{\xi}^{k},\hat{\xi}^{j}]}{4}\left[ \hat{\lambda}%
_{k}+2\left( \hat{p}_{k}+qA_{k}\right) ,\hat{\lambda}_{j}\right] .
\end{eqnarray*}
Consideration of the classical limit may be done on the states with a
definite value of $\zeta $. One can easily see that in such a limit $\hat{%
\omega}_{0}^{2}\rightarrow \omega _{0}^{2}$ and $\hat{\rho}_{1}\rightarrow
\rho $. \ The operator $\hat{\rho}_{2}$ \ is zero in classical limit, since
does not contain terms without $\hbar $. Thus, in the classical limit the
operator $\left( \hat{\Omega}^{\prime }\right) ^{2},$ and therefore $\left( 
\hat{\Omega}\right) ^{2}$ as well, corresponds to the classical quantity $%
\left. \omega ^{2}\right| _{x^{0}=\zeta \tau }$. Returning to our choice of
the operator $\hat{\Omega},$ we may say that the classical theory gives
complete information about its structure. We have to select nonclassical
parts of the operator using additional considerations. Thus, the form (\ref
{s39}) \ was selected to maintain Lorentz invariance of the results of the
quantization.

The quantum Hamiltonian (\ref{s37}) may be written in the following
block-diagonal form convenient for the further consideration, 
\begin{equation}
\hat{H}(\tau )={\rm bdiag}\left( \hat{h}(\tau ),-\hat{h}(-\tau )\right)
\;,\;\;\hat{h}(\tau )=\left. \hat{h}(x^{0})\right| _{x^{0}=\tau }\;,\;\;\hat{%
h}(x^{0})=qA_{0}+\hat{\omega}_{0}\;.  \label{s44}
\end{equation}
One can see that $\hat{h}(x^{0})$ has a form of the one-particle Dirac
Hamiltonian.

The states of the system under consideration evolute in time $\tau $ in
accordance with the Schr\"{o}dinger equation 
\begin{equation}
i\hbar \partial _{\tau }\mbox{\boldmath$\Psi$\unboldmath}(\tau )=\hat{H}%
(\tau )\mbox{\boldmath$\Psi$\unboldmath}(\tau )\;,  \label{s45}
\end{equation}
where the state vectors $\mbox{\boldmath$\Psi$\unboldmath}$ depend now
parametrically on $\tau $, 
\begin{equation}
\mbox{\boldmath$\Psi$\unboldmath}(\tau )=\left( 
\begin{array}{c}
\Psi _{+1}(\tau ,{\bf x}) \\ 
\Psi _{-1}(\tau ,{\bf x})
\end{array}
\right) \;.  \label{s46}
\end{equation}

Let us now demonstrate that the equation (\ref{s45}) is equivalent to two
Dirac equations, one for the Dirac field of the charge $q$, and another one
for the Dirac field of the charge $-q$. In accordance with our classical
interpretation we may regard $\hat{\zeta}$ as charge sign operator. Let $%
\mbox{\boldmath$\Psi$\unboldmath}_{\zeta }$ be states with a definite charge 
$\zeta q$, thus, $\hat{\zeta}\mbox{\boldmath$\Psi$\unboldmath}_{\zeta
}=\zeta \mbox{\boldmath$\Psi$\unboldmath}_{\zeta }\;,\;\;\zeta =\pm 1\;.$ It
is easily to see that states $\mbox{\boldmath$\Psi$\unboldmath}_{+1}$ with
the charge $q$ have $\Psi _{-1}=0$. In this case $\tau =x^{0}$, where $x^{0}$
is physical time. Then the equation (\ref{s45}) may be rewritten as 
\[
i\hbar \partial _{0}\Psi _{+1}(x^{0},{\bf x})=\hat{h}(x^{0})\Psi _{+1}(x^{0},%
{\bf x})\;. 
\]
Denoting $\Psi _{+1}(x^{0},{\bf x})=\psi (x)$ we get exactly the Dirac
equation for the spinor field $\psi (x)$ with charge $q$, 
\begin{equation}
\left[ \gamma ^{\mu }\left( i\hbar \partial _{\mu }-qA_{\mu }\right) -m%
\right] \psi (x)=0.  \label{s48}
\end{equation}

States $\mbox{\boldmath$\Psi$\unboldmath}_{-1}$ with charge $-q$ have $\Psi
_{+1}=0$. In this case, according to our classical interpretation, $\tau
=-x^0$, where $x^0$ is physical time. Using, for example, the standard
representation of the Dirac matrices, one can see that

\begin{equation}
\hat{h}^{c}(x^{0})=\gamma ^{2}\left( \hat{h}(x^{0})\right) ^{\ast }\gamma
^{2}=\left. \hat{h}(x^{0})\right| _{q\rightarrow -q}\;,  \label{s49}
\end{equation}
where $\hat{h}^{c}(x^{0})$ is the charge conjugated Dirac Hamiltonian. Then,
we get from the equation (\ref{s45}) 
\[
i\hbar \partial _{0}\Psi _{-1}^{\ast }(-x^{0},{\bf x})=-\gamma ^{2}\hat{h}%
^{c}(x^{0})\gamma ^{2}\Psi _{-1}^{\ast }(-x^{0},{\bf x})\;. 
\]
Denoting $\gamma ^{2}\Psi _{-1}^{\ast }(-x^{0},{\bf x})=\psi ^{c}(x)$ one
may rewrite this equation in the form of the Dirac equation for the charge
conjugated spinor field $\psi ^{c}(x)$ (that which describes particles with
the charge $-q$), 
\begin{equation}
\left[ \gamma ^{\mu }\left( i\hbar \partial _{\mu }+qA_{\mu }\right) -m%
\right] \psi ^{c}(x)=0.  \label{s50}
\end{equation}

The inner product (\ref{s33}) between two solutions of the Schr\"{o}dinger
equation (\ref{s45}) with different charges is zero. For two solutions with
charges $q$ it takes the form: 
\[
\left( \mbox{\boldmath$\Psi$\unboldmath}_{+1},\mbox{\boldmath$\Psi$%
\unboldmath}_{+1}^{\prime }\right) =\int \psi ^{+}(x)\psi (x)d{\bf x}=(\psi
,\psi ^{\prime })_{D}\;, 
\]
and is expressed via the Dirac scalar product on the $x^{0}={\rm const}$
hyperplane for the case of the charge $q$. For two solutions with charges $%
-q $ the inner product (\ref{s33}) reads: 
\[
\left( \mbox{\boldmath$\Psi$\unboldmath}_{-1},\mbox{\boldmath$\Psi$%
\unboldmath}_{-1}^{\prime }\right) =\int \psi ^{c+}(x)\psi ^{c^{\prime }}(x)d%
{\bf x}=(\psi ^{c},\psi ^{c^{\prime }})_{D}\;, 
\]
and is expressed via the Dirac scalar product for the case of the charge $-q$%
.

Let us study the eigenvalue problem for the Dirac Hamiltonian (\ref{s44}) in
a time independent external backgrounds (thus, below this Hamiltonian does
not depend on $x^{0}$): 
\begin{equation}
\hat{h}\psi ({\bf x})=\epsilon \psi ({\bf x})\;.  \label{s51}
\end{equation}
Here $\epsilon $ defines the energy spectrum of particles with the charge $q$%
. As usual, it is convenient to present the Dirac spinor in the form 
\[
\psi ({\bf x})=\left[ \gamma ^{0}\left( \epsilon -qA_{0}\right) +\gamma
^{k}\left( i\hbar \partial _{k}-qA_{k}\right) +m\right] \varphi ({\bf x})\;. 
\]
Then the function $\varphi ({\bf x})$ obeys the squared Dirac equation, 
\begin{equation}
\left[ \left( \epsilon -qA_{0}\right) ^{2}-D\right] \varphi ({\bf x}%
)=0\,,\;D=m^{2}+\left( i\hbar \partial _{k}-qA_{k}\right) ^{2}+\frac{i}{4}%
qF_{\mu \nu }[\gamma ^{\mu },\gamma ^{\nu }]_{-}\;.  \label{s52}
\end{equation}
The main features of such a spectrum in general case (for non-superstrong
potentials $A_{0}$) may be derived from the equation (\ref{s52}) repeating
the discussion presented for the scalar field. First of all, one may see
that a pair $(\varphi ,\;\epsilon )$ is a solution of the equation (\ref{s52}%
) if it obeys either the equation $\epsilon =qA_{0}+\sqrt{\varphi
^{-1}D\varphi }\;,$ or the equation $\epsilon =qA_{0}-\sqrt{\varphi
^{-1}D\varphi }\;.$ \ Let us denote via $(\varphi _{+,n},\;\epsilon _{+,n})$
solutions of the first equation, and via $(\varphi _{-,\alpha },\;\epsilon
_{-,\alpha })$ solutions of the second equation, where $n$ and $\alpha $ are
some quantum numbers which are different in general case. Thus, 
\begin{equation}
\epsilon _{+,n}=qA_{0}+\sqrt{\varphi _{+,n}^{-1}D\varphi _{+,n}}%
\;,\,\;\;\epsilon _{-,\alpha }=qA_{0}-\sqrt{\varphi _{-,\alpha
}^{-1}D\varphi _{-,\alpha }}\;.  \label{s55}
\end{equation}
One can call $\epsilon _{+,n}$ the upper branch of the energy spectrum and $%
\epsilon _{-,\alpha }$ the lower branch of the energy spectrum. We get $%
(\psi _{+,n},\;\epsilon _{+,n})$ and $(\psi _{-,\alpha },\;\epsilon
_{-,\alpha })$ solutions of the eigenvalue problem (\ref{s51}), where 
\begin{eqnarray}
\psi _{+,n}({\bf x}) &=&\left[ \gamma ^{0}\left( \epsilon
_{+,n}-qA_{0}\right) +\gamma ^{j}\left( i\hbar \partial _{j}-qA_{j}\right) +m%
\right] \varphi _{+,n}({\bf x})\;,  \nonumber \\
\psi _{-,n}({\bf x}) &=&\left[ \gamma ^{0}\left( \epsilon
_{-,n}-qA_{0}\right) +\gamma ^{j}\left( i\hbar \partial _{j}-qA_{j}\right) +m%
\right] \varphi _{-,n}({\bf x})\;.  \label{s56}
\end{eqnarray}
Square of the Dirac norm of the eigenvectors $\psi _{\varkappa ,n}$ is
positive and they may be orthonormalized as follows, 
\begin{equation}
\left( \psi _{\varkappa ,n},\psi _{\varkappa ^{\prime },n^{\prime }}\right)
_{D}=\delta _{\varkappa ,\varkappa ^{\prime }}\delta _{n,n^{\prime
}},\;\;\varkappa =\pm \;.  \label{s57}
\end{equation}

A solution of the eigenvalue problem for the charge conjugated Dirac
Hamiltonian $\hat{h}^{c}$, 
\begin{equation}
\hat{h}^{c}\psi _{\varkappa ,n}^{c}=\epsilon _{\varkappa ,n}^{c}\psi
_{\varkappa ,n}^{c}\;,  \label{s58}
\end{equation}
one can find using Eq. (\ref{s49}). Then 
\begin{equation}
\psi _{\varkappa ,n}^{c}=\gamma ^{2}\psi _{-\varkappa ,n}^{\ast
}\;,\;\;\epsilon _{\varkappa ,n}^{c}=-\epsilon _{-\varkappa ,n}\,,\;\left(
\psi _{\varkappa ,n}^{c},\psi _{\varkappa ^{\prime },n^{\prime }}^{c}\right)
=\delta _{\varkappa ,\varkappa ^{\prime }}\delta _{n,n^{\prime
}},\;\;\varkappa =\pm \;.  \label{s59}
\end{equation}
Proceeding similar to the scalar particle case in $x^{0}$-representation, we
define orthogonal each other sets $\mbox{\boldmath$\Psi$\unboldmath}%
_{+,n},\; $and $\mbox{\boldmath$\Psi$\unboldmath}_{+,\alpha }^{c}$, 
\begin{eqnarray*}
&&\hat{H}\mbox{\boldmath$\Psi$\unboldmath}_{+,n}=\epsilon _{+,n}%
\mbox{\boldmath$\Psi$\unboldmath}_{+,n},\;\;\mbox{\boldmath$\Psi$\unboldmath}%
_{+,n}=\left( 
\begin{array}{c}
\psi _{+,n}({\bf x}) \\ 
0
\end{array}
\right) \,,\;\left( \mbox{\boldmath$\Psi$\unboldmath}_{+,n},%
\mbox{\boldmath$\Psi$\unboldmath}_{+,m}\right) =\delta _{nm}\,, \\
&&\hat{H}\mbox{\boldmath$\Psi$\unboldmath}_{+,\alpha }^{c}=\epsilon
_{+,\alpha }^{c}\mbox{\boldmath$\Psi$\unboldmath}_{+,\alpha }^{c},\;\;%
\mbox{\boldmath$\Psi$\unboldmath}_{+,\alpha }^{c}=\left( 
\begin{array}{c}
0 \\ 
\psi _{+,\alpha }^{c}({\bf x})
\end{array}
\right) \,,\;\left( \mbox{\boldmath$\Psi$\unboldmath}_{+,\alpha }^{c},%
\mbox{\boldmath$\Psi$\unboldmath}_{+,\beta }^{c}\right) =\delta _{\alpha
\beta }\,,
\end{eqnarray*}
where $\hat{H}={\rm bdiag}\left( \hat{h},\;\hat{h}^{c}\right) .$ \ The sets
form a complete basis in the physical subspace ${\cal R}_{ph}^{1}$ . Now we
can see that $\hat{\Omega}$ is positive defined in the physical subspace in
accordance with positivity of classical value $\omega $. That positivity
condition helps to fix an ambiguity in the definition of $\hat{\Omega}$ .
For example, in the $\tau $-representation we could define the operator $%
\hat{\Omega}$ as follows, $\hat{\Omega}={\rm bdiag}\left( \left. \hat{\omega}%
_{0}\right| _{x^{0}=\tau },\;\pm \left. \hat{\omega}_{0}\right|
_{x^{0}=-\tau }\right) \;.$ \ We need the positivity condition to select the
minus sign in the lower block, as was done in (\ref{s39}).

To complete the consideration, as in spinless case, we examine the Dirac
quantization of the theory in question. In this case we do not need to
impose any gauge condition to the first-class constraints. We assume as
before the operator $\hat{\zeta}$ to have the eigenvalues $\zeta =\pm 1$ by
analogy with the classical theory. The equal time commutation relations for
the operators $\hat{X}^{\mu },\hat{P}_{\mu },\hat{\zeta}$, $\hat{\Xi}^{n},%
\hat{e},\hat{P}_{e},\hat{\chi},\hat{P}_{\chi }$ which correspond to the
variables $x^{\mu },p_{\mu },\zeta ,\xi ^{n},e,P_{e},\chi ,P_{\chi }$, we
define according to their Dirac brackets, with respect to second-class
constraints $\phi _{3,n}^{\left( 1\right) }$. Thus, now we get 
\begin{equation}
\ [\hat{X}^{\mu },\hat{P}_{\nu }]_{-}=i\hbar \delta _{\nu }^{\mu }\;,\;\;[%
\hat{\Xi}^{n},\hat{\Xi}^{m}]_{+}=-\frac{\hbar }{2}\eta ^{nm}\;,\;\;\hat{\zeta%
}^{2}=1\;,\;\;\left[ \hat{e},\hat{P}_{e}\right] =i\hbar \,,\;\;\left[ \hat{%
\chi},\hat{P}_{\chi }\right] _{+}=i\hbar \;,  \label{s60}
\end{equation}
whereas all other commutators (anticommutators) equals zero. Besides, we
have to keep in mind the necessity to construct an operator realization for
the first class-constraint \ $T_{2}$ from (\ref{s10}), which contains a
square root. Taking all that into account, we select as a state space one
whose elements $\mbox{\boldmath$\Psi$\unboldmath}$ are $x$-dependent
eight-component columns 
\begin{equation}
\mbox{\boldmath$\Psi$\unboldmath}=\left( 
\begin{array}{c}
\Psi _{+1}(x) \\ 
\Psi _{-1}(x)
\end{array}
\right) \;,  \label{s61}
\end{equation}
where $\Psi _{\zeta }(x),\;\zeta =\pm 1$ are four-component columns. We seek
all the operators in the block-diagonal form, 
\begin{eqnarray}
\hat{\zeta} &=&{\rm bdiag}\left( I,\;-I\right) ,\;\;\hat{X}^{\mu }=x^{\mu }%
{\bf I}\;,\;\;\;\hat{P}_{\mu }=\hat{p}_{\mu }{\bf I}\;,\;\;\hat{p}_{\mu
}=-i\hbar \partial _{\mu }\;,\;\hat{e}=e{\bf I\,,\;}\hat{P}_{e}=-i\hbar
\partial _{e}\,{\bf I},\;  \nonumber \\
\hat{\Xi}^{n} &=&{\rm bdiag}\left( \hat{\xi}^{n},\hat{\xi}^{n}\right) ,\;%
\hat{\xi}^{\mu }=\gamma ^{\mu }\hat{\xi}^{4},\;\hat{\xi}^{4}=\frac{i\hbar
^{1/2}}{2}\gamma ^{5},\;\hat{\chi}=\chi {\bf I},\;\hat{P}_{\chi }=-i\hbar
\partial _{\chi }{\bf I},  \label{s62}
\end{eqnarray}
where $I$ and ${\bf I}$ are $4\times 4$ and $8\times 8$ unit matrices
respectively, and $\;\gamma ^{5}=\gamma ^{0}...\gamma ^{3}.$The operator $%
\hat{T}_{1},$ which corresponds to the first-class constraint $T_{1}$ from (%
\ref{s9}), is selected as $\hat{T}_{1}=\hat{\Xi}^{\mu }\left( \hat{P}_{\mu
}+q\hat{A}_{\mu }\right) +m\hat{\Xi}^{4}\,,$ where $\hat{A}_{\mu }=A_{\mu }%
{\bf I}$ . The operator $\hat{T}_{2},$ which corresponds to the first-class
constraint $T_{2}$ from (\ref{s10}), is selected as $\hat{T}_{2}=\hat{P}%
_{0}+q\hat{A}_{0}+\hat{\zeta}\hat{R}\,,$ where$\;\ \hat{R}={\rm bdiag}\left( 
\hat{\omega}_{0},\;-\hat{\omega}_{0}\right) \,,\;\;\hat{\omega}_{0}=\gamma
^{0}\left[ m+\gamma ^{k}\left( \hat{p}_{k}+qA_{k}\right) \right] \;.$
Similar to the canonical quantization case, one may verify that the square $%
\hat{R}^{2}$ corresponds (in the classical limit) to the square of the
classical quantity $r,$ see the end of the Section. The state vectors (\ref
{s61}) do not depend on ''time'' $\tau $ since the Hamiltonian vanishes on
the constraints surface. The physical state vectors have to obey the
equations $\ \hat{P}_{e}\mbox{\boldmath$\Psi$\unboldmath}=0,\;\hat{P}_{\chi }%
\mbox{\boldmath$\Psi$\unboldmath}=0,\;\hat{T}_{1}\mbox{\boldmath$\Psi$%
\unboldmath}=0,\;\hat{T}_{2}\mbox{\boldmath$\Psi$\unboldmath}=0$. First two
of these conditions mean that the state vectors do not depend on $e$ and $%
\chi $ . \ Due to the bloc-diagonal form of the operators $\hat{T}$ the
second two conditions produce the following equations for the four-column $%
\Psi _{\zeta }(x),$

\begin{equation}
\hat{t}_{1}\Psi _{\zeta }(x)=0\,,\;\hat{t}_{2}\Psi _{\zeta }(x)=0\,\,,\;\
\zeta =\pm 1\,,  \label{s64}
\end{equation}
where $\hat{t}_{1}=\frac{i}{2}\hbar ^{1/2}\gamma ^{0}\;\hat{t}_{2}\gamma
^{5}\,,\;$ $\hat{t}_{2}=\hat{p}_{0}+qA_{0}+\hat{\omega}_{0}$. These
equations are consistent, the first one is a consequence of the second one.
Thus, we have in fact one equation, which may be written as follows: 
\begin{equation}
\gamma ^{0}\hat{t}_{2}\Psi _{\zeta }(x)=\left[ \gamma ^{\mu }\left( i\hbar
\partial _{\mu }-qA_{\mu }\right) -m\right] \Psi _{\zeta }(x)=0\,\,,\;\
\zeta =\pm 1\,\,.  \label{s65}
\end{equation}
Denoting $\Psi _{+1}(x)=\psi (x),$ and $\gamma ^{2}\Psi _{-1}^{\ast
}(x)=\psi ^{c}(x),$ we get two Dirac equations, one (\ref{s48})\ for the
charge $q$ and another one (\ref{s50}) for the charge $-q\;.$

Finally let us verify that $\lim_{classical}\hat{R}=r\;.$ Under the limit
sign we may replace the operator $\hat{R}$ by another one $\hat{R}^{\prime }=%
\hat{R}+\hat{\Delta},$ using the same kind of arguments, which were used in
canonical quantization case,

\[
\hat{\Delta}={\rm bdiag}\left( -\gamma ^{k}\gamma ^{5}\hat{\xi}^{0}\hat{%
\lambda}_{k},\;\gamma ^{k}\gamma ^{5}\hat{\xi}^{0}\hat{\lambda}_{k}\right)
\,,\;\;\hat{\lambda}_{k}=-4\hbar ^{-1/2}\left[ qF_{l0}\hat{\xi}^{l}\hat{\xi}%
^{0},\frac{1}{\hat{p}_{k}+qA_{k}}\right] _{+}\,.\; 
\]
Then $\left( \hat{R}^{\prime }\right) ^{2}={\rm bdiag}\left( \hat{r}^{2},\;%
\hat{r}^{2}\right) \;,\;\;\hat{r}^{2}=\hat{\omega}_{0}^{2}+\hat{\rho}_{1}+%
\hat{\rho}_{2}\,,\;\;$where 
\[
\hat{\rho}_{1}=\frac{\hbar ^{1/2}}{2i}\left[ \left( \hat{p}%
_{k}+qA_{k}\right) ,\hat{\lambda}^{k}\right] _{+},\;\hat{\rho}_{2}=-\frac{%
\hbar \hat{\lambda}_{k}^{2}}{4}-\frac{[\hat{\xi}^{k},\hat{\xi}^{j}]}{4}\left[
\hat{\lambda}_{k},\hat{\lambda}_{j}-4i\hbar ^{-1/2}\left( \hat{p}%
_{j}+qA_{j}\right) \right] . 
\]
One can easily see that in classical limit $\hat{\omega}_{0}^{2}\rightarrow
\omega _{0}^{2}$\thinspace ,\ $\hat{\rho}_{1}\rightarrow -4iqF_{l0}\,\xi
^{l}\xi ^{0},\;$ $\hat{\rho}_{2}\rightarrow 0$. \ Thus, in the classical
limit the operator $\hat{r}^{2},$ corresponds to the classical quantity $%
\left. r^{2}\right| _{\pi _{\mu }=-i\xi _{\mu }}=m^{2}+\left(
p_{k}+qA_{k}\right) ^{2}-2iqF_{\mu \nu }\xi ^{\mu }\xi ^{\nu }$.

Unfortunately, the Dirac method of the quantization gives no more
information how to proceed further with the consistent quantum theory
construction, and moreover contains principal contradictions, see discussion
in the Introduction. However, we may conclude that at least one of the main
feature of the quantum theory, its charge conjugation invariance, remains
also in the frame of the Dirac quantization.

\section{Concluding remarks}

Thus, we see that the first quantization of classical actions of spinless
and spinning particles leads to relativistic quantum mechanics which are
consistent to the same extent as corresponding quantum field theories in
one-particle sectors. Such quantum mechanics describe the corresponding
charged particles of both signs, and reproduce correctly their energy
spectra without infinite number of negative energy levels. No negative
vector norms need to be used in the corresponding Hilbert spaces.

Certainly, the relativistic quantum mechanics may not be formulated
literally in the same terms as a non-relativistic quantum mechanics. For
example, there is a problem with position and momentum operator definitions.
If one selects as such operators expressions defined by the equations (\ref
{a6}), then such operators lead state vectors out of the physical subspace.
One cannot define a positively defined probability density. All that is a
reflection of a well-known fact that it is not possible to construct
one-particle localized states in the relativistic theory. It does not depend
on the background under consideration. The problem with the momentum
operator depends on the external background, and does not exist in
translationary invariant backgrounds.

In backgrounds which violate the vacuum stability of the QFT, a more
complicated multi-particle interpretation of the quantum mechanics
constructed is also possible, which establish a connection to the QFT. Such
an interpretation will be presented in a separate publication.

Finally one ought to discuss a relation between the present results and
quantization procedure proposed earlier by Gitman and Tyutin (GT) in the
papers \cite{GitTy90a,GitTy90b}. As was already mentioned in the
Introduction the quantization there was done only for restricted classes of
external electromagnetic backgrounds, namely for constant magnetic field. \
All following attempts (see \cite{GriGr95}) to go beyond that type of
backgrounds met serious difficulties, which are not accidental. It was not
demonstrated that the quantum mechanics constructed in course of the
quantization is completely equivalent to the one-particle sector of the QFT.
In particular, one may see that quantum version of spinless particle model
does not provide right transformation properties of mean values. The
principal difference between the present approach to the quantization of RP
and the previous one is in a different understanding of the role of the
variable $\zeta $. In the papers of GT and in the following papers, which
used the same approach, they used this variable to get both branches of
solutions of Klein-Gordon equation. In course of a more deep consideration
it became clear that this aim can be achieved without the use of this
variable. One may select a special realization of the commutation relations
in the Hilbert space to get complete Klein-Gordon equation (see Sect.III).
Doing such a realization we may naturally include into the consideration
arbitrary electromagnetic and even gravitational backgrounds. Nevertheless,
the role of the variable $\zeta $ turned out to be decisive to reproduce a
consistent relativistic quantum mechanics and provide perfect equivalence
with the one-particle sector of the QFT. Due to the existence of the
variable $\zeta $ we double the Hilbert space to describe particles and
antiparticles on the same footing. Thus, we solve the problem of negative
norms and infinite number of negative energy levels. The existence of the
variable $\zeta $ makes the first and the second quantizations completely
equivalent within the one-particle sector (in cases when it may be defined
consistently). In both cases we start with an action with a given charge $q$%
, and in course of the quantization we arrive to theories which describe
particles of both charges $\pm q,$ and are $C$ -invariant. In case of the
first quantization this is achieved due to the existence and due to right
treating of the variable $\zeta $. One ought also to remark that the
requirement to maintain all classical symmetries under the coordinate
transformations and under $U(1)$ transformations allows one to realize
operator algebra without any ambiguities.

{\bf Acknowledgement} The authors are thankful to the foundations FAPESP,
and CNPq (D.M.G) for support. Besides, S.P.G. thanks the Department of
Mathematical Physics of USP for hospitality, he thanks also the Department
of Physics of UEL (Brazil) for hospitality and the Brazilian foundation
CAPES for support on the initial stage of this work.

\appendix

\section{Quantum scalar field in external backgrounds}

\subsection{Classical scalar field}

Consider here the theory of complex (charged) scalar field $\varphi (x)$
placed in a curved space-time\footnote{%
As before we use the gauge $g_{0i}=0,\;g^{00}=g_{00}^{-1}>0,\;g^{ik}g_{kj}=%
\delta _{j}^{i}$} ($g_{\mu \nu }(x)$) and interacting with an external
electromagnetic background, described by potentials $A_{\mu }(x)$. A
corresponding action for such a field theory may be written in the following
form\footnote{%
In this section we select $\hbar =c=1$}: 
\begin{equation}
S_{FT}=\int {\cal L}dx\;,\;\;{\cal L}=\sqrt{-g}\left[ \left( P_{\mu }\varphi
\right) ^{\ast }g^{\mu \nu }P_{\nu }\varphi -m^{2}\varphi ^{\ast }\varphi 
\right] ,\;\;P_{\mu }=i\partial _{\mu }-qA_{\mu }\;.  \label{b1}
\end{equation}
The corresponding Euler-Lagrange equation is covariant Klein-Gordon equation
in the background under consideration,\ 
\begin{equation}
\left[ \frac{1}{\sqrt{-g}}P_{\mu }\sqrt{-g}g^{\mu \nu }P_{\nu }-m^{2}\right]
\varphi (x)=0\;.  \label{b4}
\end{equation}

The Klein-Gordon equation was proposed by a number of authors \cite
{Klein26,Fock26,Fock26a,Gordo26}. It was shown that in a certain sense this
equation may describe particles with spin zero and the charges $(q,0,-q)$.
However, a corresponding one-particle quantum mechanics, which was
constructed to support this interpretation, contains indefinite metric and
negative energy spectrum for the positron (antiparticle) branch \cite
{FolWo50,FesVi58,Schwe61,Grein97,Weinb95}. A one-particle quantum mechanics
of particles with spin one-half and the charges $(q,0,-q)$, which was
constructed on the base of the Dirac equation \cite{Dirac28}, did not
contain indefinite metric but still cannot avoid the negative energy
spectrum for antiparticles.

The metric energy momentum tensor and the current density vector calculated
from the action (\ref{b1}) have the form 
\begin{equation}
T_{\mu \nu }=\left( P_{\mu }\varphi \right) ^{\ast }P_{\nu }\varphi +\left(
P_{\nu }\varphi \right) ^{\ast }P_{\mu }\varphi -\frac{g_{\mu \nu }}{\sqrt{-g%
}}{\cal L}\,,\;J_{\mu }=q\left[ \left( P_{\mu }\varphi \right) ^{\ast
}\varphi +\varphi ^{\ast }\left( P_{\mu }\varphi \right) \right] \;.
\label{b5}
\end{equation}
The latter obeys the continuity equation, which may be written as 
\begin{equation}
\partial _{\mu }\left( \sqrt{-g}g^{\mu \nu }J_{\nu }\right) =0\rightarrow
\partial _{0}\rho +{\rm div}{\bf j}=0,\;\rho =J_{0}g^{00}\sqrt{-g},\;{\bf j}%
=(j^{i}),\;j^{i}=g^{ik}J_{k}\sqrt{-g}.  \label{b6}
\end{equation}
Introducing the canonical momenta to the fields $\varphi ,\;\varphi ^{\ast }$%
, 
\begin{eqnarray}
&&\Pi =\frac{\partial {\cal L}}{\partial \varphi _{,0}}=i\left( P_{0}\varphi
\right) ^{\ast }g^{00}\sqrt{-g},\;\;\varphi _{,0}^{\ast }=\frac{g_{00}\Pi }{%
\sqrt{-g}}+iqA_{0}\varphi ^{\ast }\;,  \nonumber \\
&&\Pi ^{\ast }=\frac{\partial {\cal L}}{\partial \varphi _{,0}^{\ast }}%
=-i\left( P_{0}\varphi \right) g^{00}\sqrt{-g},\;\;\varphi _{,0}=\frac{%
g_{00}\Pi ^{\ast }}{\sqrt{-g}}-iqA_{0}\varphi \;,  \label{b7}
\end{eqnarray}
one may pass to Hamiltonian formulation. Calculating in this formulation
Hamiltonian, momentum, and electric charge, we get on the $x^{0}={\rm const}$
hyperplane 
\begin{eqnarray}
&&H^{FT}(x^{0})=\int T_{00}g^{00}\sqrt{-g}d{\bf x}  \nonumber \\
&=&\int \left[ \frac{g_{00}}{\sqrt{-g}}\Pi ^{\ast }\Pi -\sqrt{-g}\left(
P_{k}\varphi \right) ^{\ast }g^{kj}P_{j}\varphi +2qA_{0}{\rm Im}(\Pi \varphi
)+\sqrt{-g}m^{2}\varphi ^{\ast }\varphi \right] d{\bf x}\;,  \label{b8} \\
&&P_{i}^{FT}(x^{0})=\int T_{i0}g^{00}\sqrt{-g}d{\bf x}=\int 2{\rm Im}(\Pi
P_{i}\varphi )d{\bf x}\;,  \label{b9} \\
&&Q^{FT}=\int J_{0}g^{00}\sqrt{-g}d{\bf x}=q\int \left[ \left( P_{0}\varphi
\right) ^{\ast }\varphi +\varphi ^{\ast }\left( P_{0}\varphi \right) \right]
g^{00}\sqrt{-g}d{\bf x}=q\int \,2{\rm Im}(\Pi \varphi )d{\bf x}\;.
\label{b10}
\end{eqnarray}
The charge (\ref{b10}) does not depend on the time $x^{0}$ due to equations
of motion. Using that fact, one may introduce a conserved inner product of
two solutions of the Klein-Gordon equation 
\begin{equation}
\left( \varphi ,\varphi ^{\prime }\right) _{KG}=\int \left[ \left(
P_{0}\varphi \right) ^{\ast }\varphi ^{\prime }+\varphi ^{\ast }\left(
P_{0}\varphi ^{\prime }\right) \right] g^{00}\sqrt{-g}d{\bf x}\;.
\label{b11}
\end{equation}

\subsection{Hamiltonian form of Klein-Gordon equation}

The Klein-Gordon equation (\ref{b4}) may be rewritten in the form of a first
order in time equation (Hamiltonian form), which may be interpreted as a
Schr\"{o}dinger equation. That can be done in different ways. For example,
let us separate the time derivative part in (\ref{b4}) from the spatial one, 
\begin{equation}
i\partial _{0}\left( \sqrt{-g}g^{00}P_{0}\varphi \right) =\left[ -P_{k}\sqrt{%
-g}g^{kj}P_{j}+m^{2}\sqrt{-g}\right] \varphi +qA_{0}\left( \sqrt{-g}%
g^{00}P_{0}\varphi \right) \;.  \label{b12}
\end{equation}
Then it is easy to see that in terms of the columns 
\begin{equation}
\psi =\left( 
\begin{array}{c}
\chi \\ 
\varphi
\end{array}
\right) \;,\;\;\chi =\sqrt{-g}g^{00}P_{0}\varphi =i\Pi ^{\ast }\;
\label{b14}
\end{equation}
the equation (\ref{b12}) takes the form of the Schr\"{o}dinger equation 
\begin{equation}
i\partial _{0}\psi =\hat{h}(x^{0})\psi \;,  \label{b15}
\end{equation}
with the Hamiltonian 
\begin{eqnarray}
&&\hat{h}(x^{0})=\hat{\omega}+qA_{0},\;\;\hat{\omega}=\left( 
\begin{array}{cc}
0 & M \\ 
G & 0
\end{array}
\right) \,,\;G=\frac{g_{00}}{\sqrt{-g}}\;,  \label{b16} \\
&&M=-P_{k}\sqrt{-g}g^{kj}P_{j}+m^{2}\sqrt{-g}=-\left[ \hat{p}_{k}+qA_{k}%
\right] \sqrt{-g}g^{kj}\left[ \hat{p}_{j}+qA_{j}\right] +m^{2}\sqrt{-g},\;%
\hat{p}_{k}=-i\partial _{k}\,.  \nonumber
\end{eqnarray}
One can express the Hamiltonian (\ref{b8}), the momentum (\ref{b9}), and the
charge (\ref{b10}) in terms of the columns (\ref{b14}), 
\begin{equation}
H^{FT}(x^{0})=\int \overline{\psi }\hat{h}(x^{0})\psi d{\bf x\,}%
,\;\;P_{i}^{FT}(x^{0})=\int \overline{\psi }P_{i}\psi d{\bf x}%
\,,\;\;Q^{FT}=q\int \overline{\psi }\psi d{\bf x},\;\;\overline{\psi }=\psi
^{+}\sigma _{1}\,.  \label{b17}
\end{equation}
The continuity equation follows from (\ref{b15}) and may be written as 
\begin{equation}
\partial _{0}\rho +{\rm div}{\bf j}=0,\;\rho =\overline{\psi }\psi ,\;j^{i}=%
\frac{1}{2}g^{ik}\sqrt{-g}\left[ \overline{\psi }P_{k}+\overline{(P_{k}\psi )%
}\right] (\sigma _{1}+i\sigma _{2})\psi \;.  \label{b18}
\end{equation}
The Klein-Gordon inner product (\ref{b11}) takes then the form 
\begin{equation}
\left( \varphi ,\varphi ^{\prime }\right) _{KG}=\left( \psi ,\psi ^{\prime
}\right) =\int \overline{\psi }(x)\psi ^{\prime }(x)d{\bf x}=\int \left[
\chi ^{\ast }(x)\varphi ^{\prime }(x)+\varphi ^{\ast }(x)\chi ^{\prime }(x)%
\right] d{\bf x}\;.  \label{b19}
\end{equation}
It is easy to see that the Hamiltonian $\hat{h}(x^{0})$ is Hermitian with
respect to the inner product (\ref{b19}).

One may say that a set $\psi _{B}(x)$ (where $B$ are some quantum numbers)
of solutions of the Klein-Gordon equation (\ref{b15}) is complete if any
solution of this equation may be decomposed via the set. If this set is
orthogonal with respect to the inner product (\ref{b19}), then the
completeness relation may be written in the following form 
\begin{equation}
\left. \sum_{B}\frac{\psi _{B}(x)\overline{\psi }_{B}(y)}{(\psi _{B},\psi
_{B})}\right| _{x^{0}=y^{0}}=\delta ({\bf {x}-{y})\;.}  \label{b20}
\end{equation}
In terms of the scalar component $\varphi $ (see (\ref{b14}) this condition
reads 
\begin{equation}
\left. \sum_{B}\frac{\varphi _{B}(x)\varphi _{B}^{\ast }(y)}{(\varphi
_{B},\varphi _{B})_{KG}}\right| _{x^{0}=y^{0}}=0\,,\;\left. \sum_{B}\frac{%
\varphi _{B}(x)\sqrt{-g(y)}g^{00}(y)\left( P_{0}\varphi _{B}(y)\right)
^{\ast }}{(\varphi _{B},\varphi _{B})_{KG}}\right| _{x^{0}=y^{0}}=\delta (%
{\bf {x}-{y})\;.}  \label{b21}
\end{equation}

The Klein-Gordon equation in the common second order form (\ref{b4}) is
invariant under the operation $q\rightarrow -q,\;\varphi \rightarrow \varphi
^{c}=\varphi ^{\ast },$ which is in fact the charge conjugation operation.
That means that if $\varphi (x)$ is a wave function of a scalar particle
with a charge $q$ then $\varphi ^{c}(x)$ is a wave function for that with
the charge $-q$. For the Klein-Gordon equation in the first order form (\ref
{b15}) such an operation looks a little bit more complicated. Using the
relation (\ref{a25}), one may see that the Klein-Gordon equation in the
first order form (\ref{b15}) is invariant under the following operation $%
q\rightarrow -q,\;\psi \rightarrow \psi ^{c}=-\sigma _{3}\psi ^{\ast },$ so
that 
\begin{equation}
i\partial _{0}\psi ^{c}=\hat{h}^{c}(x^{0})\psi ^{c},\;\;\hat{h}%
^{c}(x^{0})=\left. \hat{h}(x^{0})\right| _{q\rightarrow -q}=-\left[ \sigma
_{3}\hat{h}(x^{0})\sigma _{3}\right] ^{\ast }\;.  \label{b24}
\end{equation}
Thus defined charge conjugation for two columns (\ref{b14}) is matched with
the charge conjugation for scalar wave functions.

\subsection{Solutions and spectrum of Klein-Gordon equation}

Let us study the eigenvalue problem for the Hamiltonian (\ref{b16}) in time
independent external backgrounds (thus, below this Hamiltonian does not
depend on $x^{0}$): 
\begin{equation}
\hat{h}\psi ({\bf x})=\epsilon \psi ({\bf x})\;,\;\;\psi ({\bf x})=\left( 
\begin{array}{c}
\chi ({\bf x}) \\ 
\varphi ({\bf x})
\end{array}
\right) \;.  \label{b25}
\end{equation}
Being written in components, the equation (\ref{b25}) takes the form: 
\begin{equation}
qA_{0}\chi +M\varphi =\epsilon \chi \,,\;\;qA_{0}\varphi +G\chi =\epsilon
\varphi \;.  \label{b26}
\end{equation}
The system (\ref{b26}) results in the following equation for $\varphi $: 
\begin{equation}
GM\varphi =[\epsilon -qA_{0}]^{2}\varphi \;\Rightarrow \left[ \left(
\epsilon -qA_{0}\right) ^{2}g^{00}+\frac{1}{\sqrt{-g}}P_{k}\sqrt{-g}%
g^{kj}P_{j}-m^{2}\right] \varphi =0\;.  \label{b27}
\end{equation}
If we make the substitution $\varphi (x)=\exp [-i\epsilon x^{0}]\varphi (%
{\bf x})$ in the Klein-Gordon equation (\ref{b4}) we arrive just to the
equation (\ref{b27}). Thus, $\epsilon $ defines the energy spectrum of the
Klein-Gordon equation for particles with the charge $q$. Such a spectrum is
well known for free background and for special exact solvable cases of
external electromagnetic and gravitational fields \cite
{GreMuR85,BagGi90,FraGiS91,BirDa82,GriMaM88}. The main features of such a
spectrum in general case (for non-superstrong potentials $A^{0}$) may be
derived from the equation (\ref{b27}). First of all, one may see that a pair 
$(\varphi ,\;\epsilon )$ is a solution of the equation (\ref{b27}) if it
obeys either the equation $\epsilon =qA_{0}+\sqrt{\varphi ^{-1}GM\varphi }%
\;, $ or the equation $\epsilon =qA_{0}-\sqrt{\varphi ^{-1}GM\varphi }\;.$ \
Let us denote via $(\varphi _{+,n},\;\epsilon _{+,n})$ solutions of the
first equation, and via $(\varphi _{-,\alpha },\;\epsilon _{-,\alpha })$
solutions of the second equation, where $n$ and $\alpha $ are some quantum
numbers which are different in general case. Thus, 
\begin{equation}
\epsilon _{+,n}=qA_{0}+\sqrt{\varphi _{+,n}^{-1}GM\varphi _{+,n}}%
\;,\;\;\epsilon _{-,\alpha }=qA_{0}-\sqrt{\varphi _{-,\alpha }^{-1}GM\varphi
_{-,\alpha }}\;.  \label{b31}
\end{equation}
It is clear that 
\begin{equation}
\epsilon _{+,n}-\epsilon _{-,\alpha }=\sqrt{\varphi _{+,n}^{-1}GM\varphi
_{+,n}}+\sqrt{\varphi _{-,\alpha }^{-1}GM\varphi _{-,\alpha }}\;>0\;.
\label{b32}
\end{equation}
Thus, one can call $\epsilon _{+,n}$ the upper branch of the energy spectrum
and $\epsilon _{-,\alpha }$ the lower branch of the energy spectrum. In the
presence of the potential $A_{0}$ they may be essentially nonsymmetric, as
an example one can remember the energy spectrum in Coulomb field, where (for
an attractive Coulomb potential for the charge $q$) the upper branch
contains both discrete and continuous parts of energy levels and the lower
branch contains only continuous levels, see Fig.1.

\bigskip 
\unitlength=0.8pt 
\begin{picture}(220,410)
\put(0,200){\line(1,0){300}}
\put(110,0){\vector(0,1){400}}
\put(105,405){{\large $\varepsilon$}}
\multiput(105,265)(0,4){30}{\line(1,0){10}}
\multiput(105,135)(0,-4){30}{\line(1,0){10}}
\put(62,258){\phantom{$-$}$mc^2$}
\put(62,137){$-mc^2$}
{\thicklines \put(100,140){\line(1,0){20}}
             \put(100,225){\line(1,0){20}}
             \put(100,240){\line(1,0){20}}
             \put(100,249){\line(1,0){20}}
             \put(100,256){\line(1,0){20}}
             \put(100,260){\line(1,0){20}} }
\put(125,73) { $\left. {\rule{0pt}{56pt}} \right \}$ }
\put(125,302){ $\left. {\rule{0pt}{69pt}} \right \}$ }
\put(144,73){lower branch:} 
\put(144,55){$\varepsilon_{-,\a},\;\psi_{-,\a}$}
\put(144,302){upper branch:} 
\put(144,284){$\varepsilon_{+,n},\;\psi_{+,n}$}
\put(75,0){a)}
\put(240,0){ 
 \begin{picture}(220,410)
 \put(0,200){\line(1,0){220}}
 \put(110,0){\vector(0,1){400}}
 \put(105,405){{\large $\varepsilon^{c}$}}
 \multiput(105,265)(0,4){30}{\line(1,0){10}}
 \multiput(105,135)(0,-4){30}{\line(1,0){10}}
 \put(62,258){\phantom{$-$}$mc^2$}
 \put(62,137){$-mc^2$}
 {\thicklines \put(100,140){\line(1,0){20}}
             \put(100,144){\line(1,0){20}}
             \put(100,151){\line(1,0){20}}
             \put(100,160){\line(1,0){20}}
             \put(100,175){\line(1,0){20}}
             \put(100,260){\line(1,0){20}} }
 \put(125,90)  { $\left. {\rule{0pt}{69pt}} \right \}$ }
 \put(125,320){ $\left. {\rule{0pt}{56pt}} \right \}$ }
 \put(144,90){lower branch:} 
 \put(144,72){$\varepsilon_{-,n}^{c},\; \psi^c_{-,n}$}
 \put(144,320){upper branch:} 
 \put(144,302){$\varepsilon_{+,\a}^{c},\; \psi^c_{+,\a}$}
 \put(75,0){b)}
 \end{picture}}

\end{picture}
\bigskip

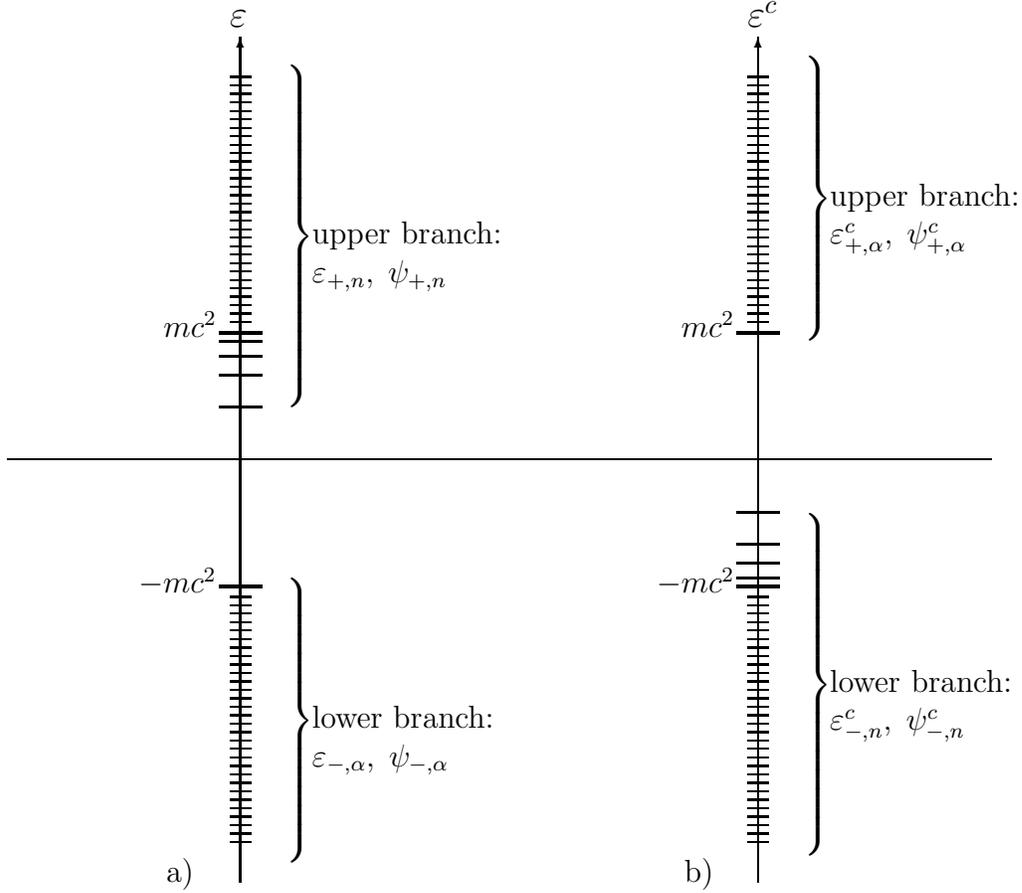
{\bf Fig.1.}{\ Energy spectra of Klein-Gordon particles with a charge $q$
and $-q$; a) - spectrum of $\hat h$, b) - spectrum of $\hat h^c$.}

\bigskip

In the absence of the potential $A_0$, one can always select equal quantum
numbers for both branches, thus, the total spectrum becomes symmetric, 
\[
\epsilon_{\pm,n}=\pm \sqrt{\varphi^{-1}_{\pm,n} GM\varphi}_{\pm,n}=
\mp\,\epsilon_{\mp,n}\;. 
\]
Further, even in general case when $A_0$ is not zero, we are going to use
sometimes the same index $n$ to label quantum numbers both for upper and
lower branches to simplify equations, hoping that it does not lead to a
misunderstanding for those readers who keeps in mind the above explanations.

We may express the functions $\chi $ from the equation (\ref{b26}), 
\[
\chi _{\varkappa ,n}=g^{00}\sqrt{-g}\left( \epsilon _{\varkappa
,n}-qA_{0}\right) \varphi _{\varkappa ,n},\;\;\varkappa =\pm \;.
\]
Thus, 
\begin{equation}
\hat{h}\psi _{\varkappa ,n}=\epsilon _{\varkappa ,n}\psi _{\varkappa
,n},\;\psi _{\varkappa ,n}=\left( 
\begin{array}{c}
g^{00}\sqrt{-g}\left( \epsilon _{\varkappa ,n}-qA_{0}\right) \varphi
_{\varkappa ,n} \\ 
\varphi _{\varkappa ,n}
\end{array}
\right) \;.  \label{b33}
\end{equation}
Calculating square of the norm of the eigenvectors $\psi _{\varkappa ,n}$,
using the inner product (\ref{b19}), we find 
\begin{equation}
\left( \psi _{\varkappa ,n},\psi _{\varkappa ,n}\right) =2\int \left(
\epsilon _{\varkappa ,n}-qA_{0}\right) |\varphi _{\varkappa ,n}|^{2}g^{00}%
\sqrt{-g}d{\bf x}\;.  \label{b34}
\end{equation}
Taking into account the positivity of $g^{00}$ and the relation ${\rm sign}%
\left( \epsilon _{\varkappa ,n}-qA_{0}\right) =\varkappa \;,$ which follows
from (\ref{b31}), we can see that ${\rm sign}\left( \psi _{\varkappa
,n},\psi _{\varkappa ,n}\right) =\varkappa \;.$ Since the Hamiltonian $%
\hat{h}$ is Hermitian with respect to the inner product (\ref{b19}), we get
for the normalized eigenvectors $\psi _{\varkappa ,n}$ the following
orthonormality conditions 
\begin{equation}
\left( \psi _{\varkappa ,n},\psi _{\varkappa ^{\prime },n^{\prime }}\right)
=\varkappa \delta _{\varkappa ,\varkappa ^{\prime }}\delta _{n,n^{\prime
}}\,.  \label{b36}
\end{equation}

The set $\psi _{\varkappa ,n}$ is complete in the space of two columns
dependent on ${\bf x}$. An explicit form of the completeness relation may be
written if one takes into account equations (\ref{b20}) and (\ref{b36}), 
\begin{equation}
\sum_{n}\left[ \psi _{+,n}({\bf {x})\overline{\psi }_{+,n}({y})-\psi _{-,n}({%
x}){\overline{\psi }}_{-,n}({y})}\right] =\delta ({\bf {x}-{y})\;.}
\label{b37}
\end{equation}

One can easily see that the equation (\ref{b27}) retains his form under the
following substitution $\epsilon \rightarrow -\epsilon ,\;q\rightarrow
-q,\;\varphi \rightarrow \varphi ^{\ast }.$ That means that that the energy
spectrum $\epsilon ^{c}$ of the Klein-Gordon equation for the charge $-q$ is
related to the energy spectrum $\epsilon $ of the Klein-Gordon equation for
the charge $q$ by the relation $\epsilon ^{c}=-\epsilon \;.$

Using similar consideration, it is possible to present a solution of the
eigenvalue problem for the charge conjugated Hamiltonian $\hat{h}^{c}$, see
Eq. (\ref{b24}). In fact, the result may be derived from (\ref{b33}) by use (%
\ref{b24}). It reads (see Fig. 1): 
\begin{equation}
\hat{h}^{c}\psi _{\varkappa ,n}^{c}=\epsilon _{\varkappa ,n}^{c}\psi
_{\varkappa ,n}^{c}\,,\;\psi _{\varkappa ,n}^{c}=-\sigma _{3}\psi
_{-\varkappa ,n}^{\ast }\,,\;\epsilon _{\varkappa ,n}^{c}=-\epsilon
_{-\varkappa ,n},\;\left( \psi _{k,n}^{c}\,,\psi _{\varkappa ^{\prime
},n^{\prime }}^{c}\right) =\varkappa \delta _{\varkappa ,\varkappa ^{\prime
}}\delta _{n,n^{\prime }}\,.  \label{b39}
\end{equation}

It is easy to see that the charge conjugated solutions $\psi _{\varkappa
,n}^{c}$ obey the same orthonormality conditions (\ref{b36}) and the
completeness relation (\ref{b37}). The latter being written in terms of $%
\psi $ and $\psi ^{c}$ takes the form 
\begin{equation}
\sum_{n}\left[ \psi _{+,n}({\bf {x})\overline{\psi }_{+,n}({y})+\sigma
_{3}\psi _{+,n}^{c\ast }({x})\overline{\psi _{+,n}^{c}}^{\ast }({y})\sigma
_{3}}\right] =\delta ({\bf {x}-{y})\;.}  \label{b40}
\end{equation}
It involves now only positive energy solutions for particles and
antiparticles.

Time dependent set of solutions $\psi _{\varkappa ,n}(x)$ of the
Klein-Gordon equation (\ref{b15}), which is related to the stationary set of
eigenvectors $\psi _{\varkappa ,n}({\bf x})$, reads as follows: 
\begin{equation}
\psi _{\varkappa ,n}(x)=\exp \{-i\varepsilon _{\varkappa ,n}x^{0}\}\psi
_{\varkappa ,n}({\bf x}).  \label{b41}
\end{equation}
It is complete and obeys the orthonormality conditions (\ref{b36}).

\subsection{Quantized scalar field}

In course of the quantization (second quantization) the fields $\varphi $
and $\Pi $ become Heisenberg operators with equal-time commutation relations 
$[\hat{\varphi}(x),\hat{\Pi}(y)]_{x^{0}=y^{0}}=i\delta ({\bf x}-{\bf y})\;,$
which imply the following commutation relations for the Heisenberg operators 
$\hat{\psi}(x)$ (operator columns of the form (\ref{b14})) and $\hat{\psi}%
^{c}=-(\hat{\psi}^{+}\sigma _{3})^{T}$: 
\begin{equation}
\lbrack \hat{\psi}(x),\hat{\overline{\psi }}(y)]_{x^{0}=y^{0}}=[\hat{\psi}%
^{c}(x),\hat{\overline{\psi }}^{c}(y)]_{x^{0}=y^{0}}=\delta ({\bf x}-{\bf y}%
)\;.  \label{b47}
\end{equation}
Equations of motion for the operators $\hat{\psi}$ and $\hat{\psi}^{c}$ have
the form 
\begin{equation}
i\partial _{0}\hat{\psi}(x)=[\hat{\psi}(x),\hat{H}^{FT}(x^{0})]=\hat{h}%
(x^{0})\hat{\psi}(x)\,,\;i\partial _{0}\hat{\psi}^{c}(x)=\hat{h}^{c}(x^{0})%
\hat{\psi}^{c}(x)\;,  \label{b49}
\end{equation}
where $\hat{h}(x^{0})$ and $\hat{h}^{c}(x^{0})$ are defined by (\ref{b16})
and (\ref{b24}) respectively. The first equation (\ref{b49}) implies the
Klein-Gordon equation (\ref{b4}) for the Heisenberg field $\hat{\varphi}(x)$.

In external backgrounds, which do not create particles from the vacuum, one
may define subspaces (in the Hilbert space of the quantum theory of a field)
with definite numbers of particles invariant under the evolution \cite
{BirDa82,GreMuR85,GriMaM88,Fulli89,FraGiS91}. An important example of the
above backgrounds are nonsingular time independent external backgrounds%
\footnote{%
As examples of singular time independent external backgrounds one may
mention supercritical Coulomb fields, and electric fields in time
independent gauges with infinitely growing potentials on the space infinity}%
. Let us consider below such kind of backgrounds to simplify the
demonstration. A generalization to arbitrary backgrounds, in which the
vacuum remains stable, looks similar \cite{FraGiS91}.

One may decompose the Heisenberg operator $\hat{\psi}(x)$ in the complete
set (\ref{b41}), 
\begin{equation}
\hat{\psi}(x)=\sum_{n}\left[ a_{n}\psi _{+,n}(x)+b_{n}^{+}\psi _{-,n}(x)%
\right] \;.  \label{b50}
\end{equation}
It follows from the commutation relations (\ref{b47}) and from the
orthonormality relations (\ref{b36}) that $%
[a_{n},a_{m}^{+}]=[b_{n},b_{m}^{+}]=\delta
_{nm},\;[a_{n},a_{m}]=[b_{n},b_{m}]=0\;.$ Thus, we get two sets of
annihilation and creation operators $a_{n},a_{n}^{+}$ and $b_{n},b_{n}^{+}$,
which may be interpreted as ones of particles with a charge $q$ and
antiparticles with a charge $-q$. Indeed, the quantum Hamiltonian and the
operator of the charge, which may be constructed from the expressions (\ref
{b17}), have the following diagonal form in terms of such creation and
annihilation operators 
\begin{eqnarray}
&&\hat{H}^{FT}=\hat{H}_{R}^{FT}+E_{0},\;\hat{H}_{R}^{FT}=\sum_{n}\left[
\epsilon _{+,n}a_{n}^{+}a_{n}-\epsilon _{-,n}b_{n}^{+}b_{n}\right] =\sum_{n}%
\left[ \epsilon _{+,n}a_{n}^{+}a_{n}+\epsilon _{+,n}^{c}b_{n}^{+}b_{n}\right]
\;,  \nonumber \\
&&\hat{Q}^{FT}=q\sum_{n}\left[ a_{n}^{+}a_{n}-b_{n}^{+}b_{n}\right] \;,
\label{b52}
\end{eqnarray}
where $E_{0}=-\sum_{n}\epsilon _{-,n}=\sum_{n}\epsilon _{+,n}^{c}$ is an
infinite constant, and $\hat{H}_{R}^{FT}$ is a renormalized Hamiltonian,
namely the latter is selected as the energy operator.

The Hilbert space ${\cal R}^{FT}$ of the quantum field theory may be
constructed in the backgrounds under consideration as a Fock space. One
defines the vacuum state $|0>$ as a zero vector for all the annihilation
operators $a_{n}|0>=b_{n}|0>=0\;.$ \ The energy of such defined vacuum is
zero. A complete basis may be constructed by means of the action of the
creation operators on the vacuum, $a_{n_{1}}^{+}\ldots
a_{n_{A}}^{+}b_{\alpha _{1}}^{+}\ldots b_{\alpha
_{B}}^{+}|0>,\;\;A,B=0,1,...\;.$ At a fixed $A$ and $B$ the basis vectors
describe states with $A$ particles and $B$ antiparticles with given quantum
numbers respectively. A state vector of the quantum field theory in a given
time instant $x^{0}$ we denote as $|\mbox{\boldmath$\Psi$\unboldmath}%
(x^{0})> $. It evolutes with the time $x^{0}$ according to the
Schr\"{o}dinger equation with the renormalized Hamiltonian $\hat{H}_{R}^{FT}$%
, 
\begin{equation}
i\partial _{0}|\mbox{\boldmath$\Psi$\unboldmath}(x^{0})>=\hat{H}_{R}^{FT}|%
\mbox{\boldmath$\Psi$\unboldmath}(x^{0})>\;.  \label{b55}
\end{equation}
In the time independent background under consideration each subspace ${\cal R%
}_{AB}^{FT}$ of state vectors with the given number of particles $A$ and
antiparticles $B$ is invariant under the time evolution, since the
Hamiltonian $\hat{H}_{FT}$ commutes with number of particles operator $\hat{N%
}$, 
\begin{equation}
\hat{N}=\sum_{n}\left[ a_{n}^{+}a_{n}+b_{n}^{+}b_{n}\right] \;.  \label{b57}
\end{equation}



\end{document}